\documentclass[traditabstract]{aa} 
\usepackage{txfonts}
\usepackage{graphicx}
\usepackage{longtable}

\begin{document}

\title{The MUCHFUSS project -- Searching for hot subdwarf binaries with massive unseen companions}
\subtitle{
Survey, target selection and atmospheric parameters
\thanks{Based on observations at the Paranal Observatory of the European 
Southern Observatory for programme number 081.D-0819. Based on observations at the La Silla Observatory of the European 
Southern Observatory for programme number 082.D-0649. Based on observations collected at 
  the Centro Astron\'omico Hispano Alem\'an (CAHA) at Calar Alto, operated 
  jointly by the Max-Planck Institut f\"ur Astronomie and the Instituto de
   Astrof\'isica de Andaluc\'ia (CSIC). Based on observations with the William Herschel Telescope operated by the Isaac Newton Group at the Observatorio del Roque de los Muchachos of the Instituto de Astrofisica de Canarias on the island of La Palma, Spain.}
}

\author{S. Geier \inst{1}
   \and H. Hirsch \inst{1}
   \and A. Tillich \inst{1}
   \and P. F. L. Maxted \inst{2}
   \and S. J. Bentley \inst{2}
   \and R. H. \O stensen \inst{3}
   \and U. Heber \inst{1}
   \and B. T. G\"ansicke \inst{4}
   \and T. R. Marsh \inst{4}
   \and R. Napiwotzki \inst{5}
   \and B. N. Barlow \inst{6}
   \and S. J. O'Toole \inst{1,7}}

\offprints{S.\,Geier,\\ \email{geier@sternwarte.uni-erlangen.de}}

\institute{Dr. Karl Remeis-Observatory \& ECAP, Astronomical Institute, Friedrich-Alexander University Erlangen-Nuremberg, Sternwartstr.~7, D 96049 Bamberg, Germany
\and Astrophysics Group, Keele University, Staffordshire, ST5 5BG, UK
\and Institute of Astronomy, K.U.Leuven, Celestijnenlaan 200D, B-3001 Heverlee, Belgium
\and Department of Physics, University of Warwick, Conventry CV4 7AL, UK
\and Centre of Astrophysics Research, University of Hertfordshire, College
  Lane, Hatfield AL10 9AB, UK
\and Department of Physics and Astronomy, University of North Carolina, Chapel Hill, NC 27599-3255, USA
\and Australian Astronomical Observatory, PO Box 296, Epping, NSW, 1710, Australia}

\date{Received \ Accepted}

\abstract{
The project Massive Unseen Companions to Hot Faint Underluminous Stars from SDSS (MUCHFUSS) aims at finding sdBs with compact companions like supermassive white dwarfs ($M>1.0\,{\rm M_{\odot}}$), neutron stars or black holes. The existence of such systems is predicted by binary evolution theory and recent discoveries indicate that they are likely to exist in our Galaxy.\\
A determination of the orbital parameters is sufficient to put a lower limit on the companion mass by calculating the binary mass function. If this lower limit exceeds the Chandrasekhar mass and no sign of a companion is visible in the spectra, the existence of a massive compact companion is proven without the need for any additional assumptions. We identified about $1100$ hot subdwarf stars from the SDSS by colour selection and visual inspection of their spectra. Stars with high velocities have been reobserved and individual SDSS spectra have been analysed. In total $127$ radial velocity variable subdwarfs have been discovered. Binaries with high RV shifts and binaries with moderate shifts within short timespans have the highest probability of hosting massive compact companions. Atmospheric parameters of $69$ hot subdwarfs in these binary systems have been determined by means of a quantitative spectral analysis. The atmospheric parameter distribution of the selected sample does not differ from previously studied samples of hot subdwarfs. The systems are considered the best candidates to search for massive compact companions by follow-up time resolved spectroscopy. 

}

\maketitle

\section{Introduction \label{sec:intro}}

Subuminous B stars (sdBs) are core helium-burning stars with very thin hydrogen envelopes and masses around $0.5\,{\rm M_{\odot}}$ 
(Heber \cite{heber86}). A large fraction of the sdB stars are members of short period binaries (Maxted et al. \cite{maxted01};  Napiwotzki et al. \cite{napiwotzki04a}). After the discovery of close binary subdwarfs, several studies aimed at determining the fraction of hot subdwarfs residing in such systems. Samples of hot subdwarfs have been checked for radial velocity (RV) variations. The binary fraction has been determined to range from $39\,\%$ to $78\,\%$ (e.g. Maxted et al. \cite{maxted01};  Napiwotzki et al. \cite{napiwotzki04a}). Several studies were undertaken to determine the orbital parameters of subdwarf binaries (e.g. Edelmann et al. \cite{edelmann05}; Morales-Rueda et al. \cite{morales03a}). The orbital periods range from $0.07$ to $>10\,{\rm d}$ with a peak at $0.5-1.0\,{\rm d}$.  

\begin{figure}[t!]
	\resizebox{\hsize}{!}{\includegraphics{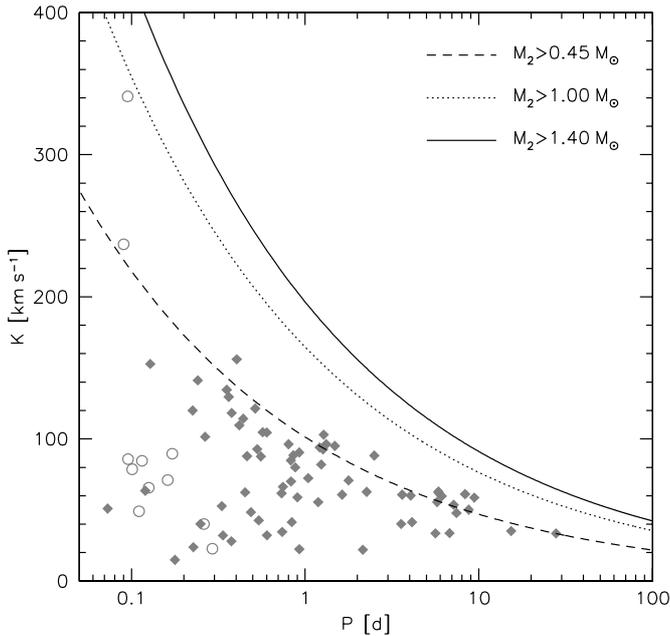}}
	\caption{The RV semiamplitudes of all known sdB binaries with spectroscopic solutions plotted against their orbital periods (see Table~\ref{tab:orbitslit}). Binaries which have initially been discovered in photometric surveys due to indicative features in their light curves (eclipses, reflection effects, ellipsoidal variations) are marked with open circles. Binaries discovered by detection of RV variation from time resolved spectroscopy are marked with filled diamonds. The dashed, dotted and solid lines mark the regions to the right where the minimum companion masses derived from the binary mass function (assuming $0.47\,{\rm M_{\odot}}$ for the sdBs) exceed $0.45\,{\rm M_{\odot}}$, $1.00\,{\rm M_{\odot}}$ and $1.40\,{\rm M_{\odot}}$. The two post-RGB objects in the sample have been excluded, because their primary masses are much lower.}
\label{periodK_lit}
\end{figure}

For close binary sdBs common envelope ejection is the most probable formation channel (Han et al. \cite{han02,han03}). In this scenario two main sequence stars of different masses evolve in a binary system. The more massive one will reach the red giant phase first and fill its Roche lobe near the tip of the red-giant branch. If the mass transfer to the companion is dynamically unstable, a common envelope is formed. Due to friction the two stellar cores lose orbital energy, which is deposited within the envelope and leads to a shortening of the binary period. Eventually the common envelope is ejected and a close binary system is formed, which contains a core helium-burning sdB and a main sequence companion. A binary consisting of a main sequence star and a white dwarf may evolve to a close binary sdB with a white dwarf companion in a similar way. Only in very special and hence rare cases tight constraints can be put on the nature of the companions, that is if the systems are eclipsing or show other indicative features in their light curves (see the catalogue of Ritter \& Kolb \cite{ritter03} and references therein). 

Subdwarf binaries with massive WD companions turned out to be candidates for SN Ia progenitors because these systems lose angular momentum due to the emission of gravitational waves and start mass transfer. The mass transfer, or the subsequent merger of the system, may cause the WD to approach the Chandrasekhar limit, ignite carbon under degenerate conditions and explode as a SN Ia (Webbink \cite{webbink84}; Iben \& Tutukov \cite{iben84}). One of the best known candidate system for this double degenerate merger scenario is the sdB+WD binary KPD\,1930$+$2752 (Maxted et al. \cite{maxted00a}; Geier et al. \cite{geier07}). Mereghetti et al. (\cite{mereghetti09}) showed that in the X-ray binary HD\,49798 a massive ($>1.2\,{\rm M_{\odot}}$) white dwarf accretes matter from a closely orbiting subdwarf O companion. The predicted amount of accreted material is sufficient for the WD to reach the Chandrasekhar limit. This makes HD\,49798 another candidate SN\,Ia progenitor, should the companion be a C/O white dwarf (Wang et al. \cite{wang09}). SN~Ia play a key role in the study of cosmic evolution. They are utilised as standard candles for determining the cosmological parameters (e.g. Riess et al. \cite{riess98}; 
Leibundgut \cite{leibundgut01}; Perlmutter et al. \cite{perlmutter99}). Most recently Perets et al. (\cite{perets10}) showed that helium accretion onto a white dwarf may be responsible for a subclass of faint and calcium-rich SN Ib events.

Due to the tidal influence of the companion in close binary systems, the rotation of the primary\footnote{The more massive component of a binary is usually defined as the primary. However, in most close sdB binaries with unseen companions the masses are unknown and it is not possible to decide a priori which component is the most massive one. For this reason we call the visible sdB component of the binaries the primary throughout this paper.} becomes  synchronised to its orbital motion. In this case it is possible to constrain the mass of the companion, if  mass, projected rotational velocity and surface gravity of the sdB are known. Geier et al. (\cite{geier08}, \cite{geier10a}, \cite{geier10b}) analysed high resolution spectra of 41 sdB stars in close binaries, half of all systems with known orbital parameters. In 31 cases, the mass and nature of the unseen companions could be constrained. While most of the derived companion masses were consistent with either late main sequence stars or white dwarfs, the compact companions of some sdBs may be either massive white dwarfs, neutron stars (NS) or stellar mass black holes (BH). However, Geier et al. (\cite{geier10b}) also showed that the assumption of orbital synchronisation in close sdB binaries is not always justified and that their sample suffers from huge selection effects. 

The existence of sdB+NS/BH systems is predicted by binary evolution theory (Podsiadlowski et al. \cite{podsi02}; Pfahl et al. \cite{pfahl03}). The formation channel includes two phases of unstable mass transfer and one supernova explosion. The predicted fraction of sdB+NS/BH systems ranges from about $1\%$ to $2\%$ of the close sdB binaries (Geier et al. \cite{geier10b}; Yungelson \& Tutukov \cite{yungelson05}; Nelemans \cite{nelemans10}).

\begin{figure*}[t!]
\begin{center}
        \resizebox{8.5cm}{!}{\includegraphics{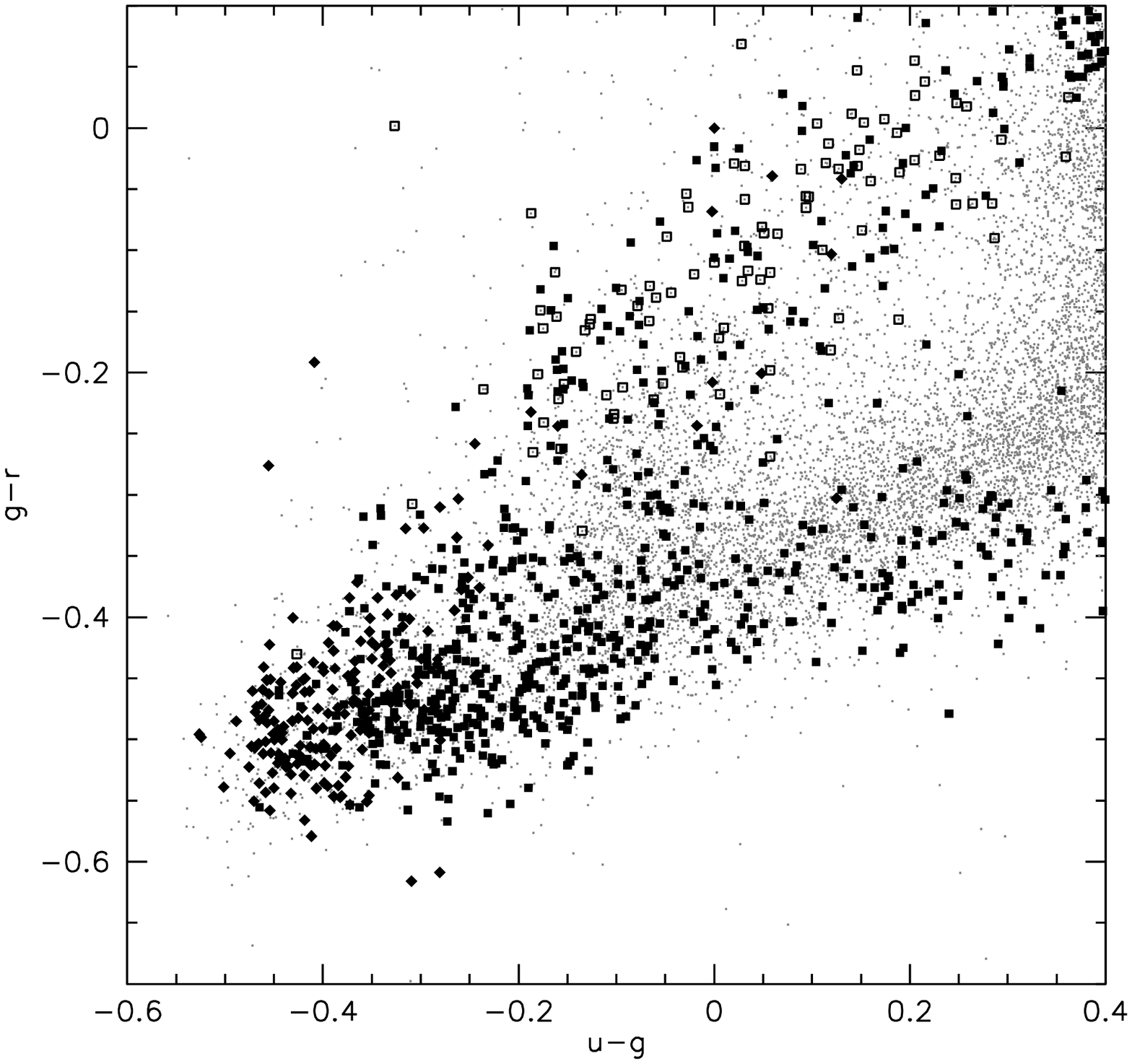}}
	\resizebox{8.5cm}{!}{\includegraphics{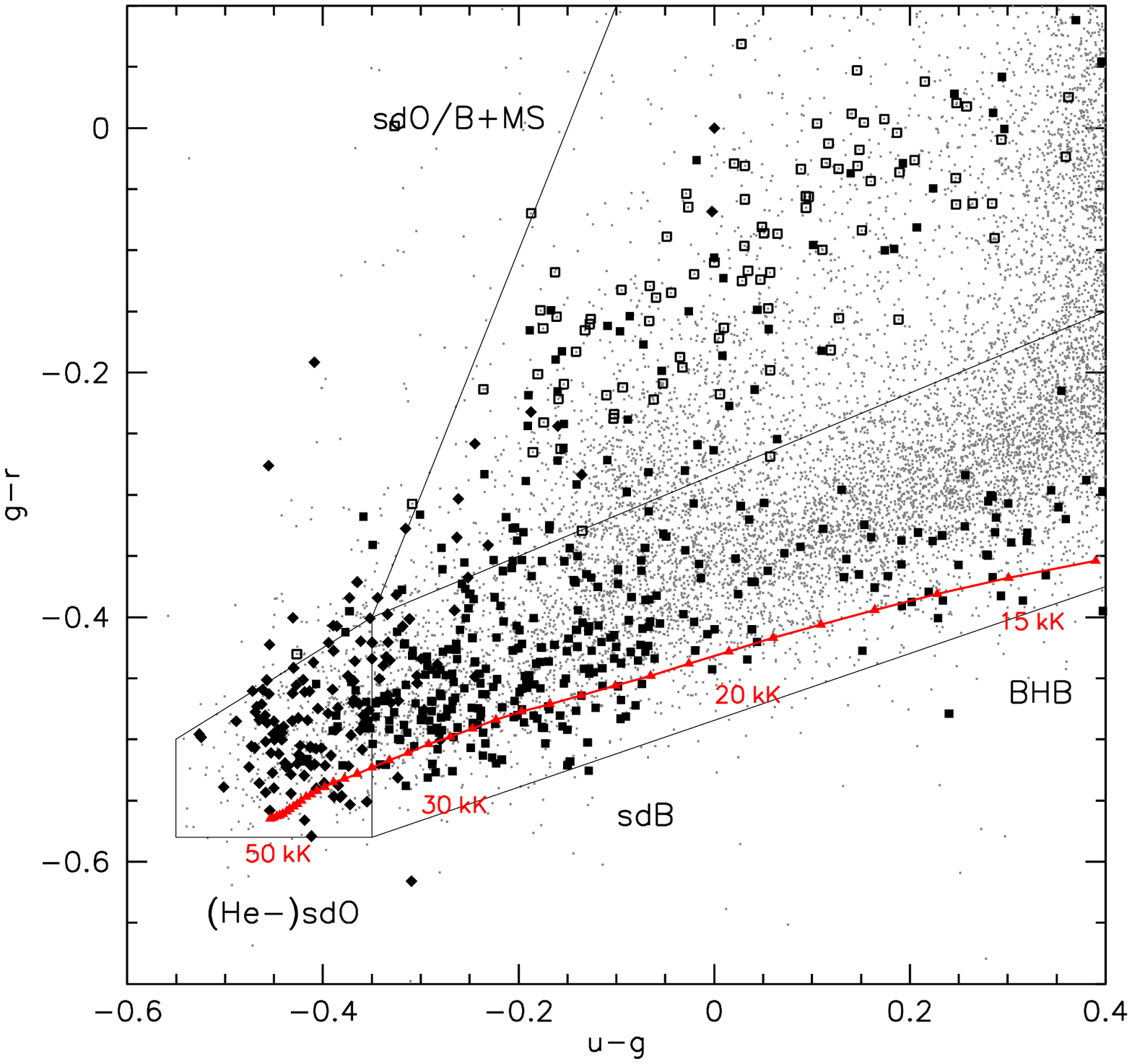}}
\end{center}
\caption{{\it Left panel.} SDSS $g-r$-colours plotted against $u-g$ of all stars. The grey dots mark all stellar objects with spectra available in the SDSS database. Most of them are classified as DA white dwarfs. The solid diamonds mark (He-)sdO stars, the solid squares sdB and sdOB stars. Open squares mark hot subdwarfs with main sequence companions visible in the spectra. Most of these objects are white dwarfs of DA type. {\it Right panel.} Only subdwarfs with $g<18\,{\rm mag}$ are plotted. The sequence of composite objects is clearly separated from the single-lined stars. Synthetic colours from Castelli \& Kurucz (\cite{castelli03}) for stars with temperatures ranging from $14\,000\,{\rm K}$ to $50\,000\,{\rm K}$ ($\log{g}=5.0$) are marked with upward triangles and connected. The stepsize of the colour grid is $1000\,{\rm K}$. The labels mark models of certain temperatures.}
\label{colours_mag}
\end{figure*}

\section{Project overview}\label{s:much}

The work of Geier et al. (\cite{geier10b}) indicates that a population of non-interacting binaries with massive compact companions may be present in our Galaxy. The candidate sdB+NS/BH binaries have low orbital inclinations ($15-30^{\rm \circ}$, Geier et al. \cite{geier10b}). High inclination systems must exist as well. A determination of the orbital parameters allows one to put a lower limit to the companion mass by calculating the binary mass function. 

\begin{equation}
\label{equation-mass-function}
 f_{\rm m} = \frac{M_{\rm comp}^3 \sin^3i}{(M_{\rm comp} +
   M_{\rm sdB})^2} = \frac{P K^3}{2 \pi G}
\end{equation}

The RV semi-amplitude $K$ and the period $P$ can be derived from the RV curve; the sdB mass $M_{\rm sdB}$, the companion mass $M_{\rm comp}$ and the inclination angle $i$ remain free parameters. We adopt $M_{\rm sdB}=0.47\,{\rm M_{\odot}}$ and $i<90^{\rm \circ}$ to derive a lower limit for the companion mass. Depending on this minimum mass a qualitative classification of the companions' nature is possible in certain cases. For minimum companion masses lower than $0.45\,{\rm M_{\odot}}$ a main sequence companion can not be excluded because its luminosity would be too low to be detectable in the spectra (Lisker et al. \cite{lisker05}). If the minimum companion mass exceeds $0.45\,{\rm M_{\odot}}$ and no spectral signatures of the companion are visible, it must be a compact object. If it exceeds the Chandrasekhar mass and no sign of a companion is visible in the spectra, the existence of a massive compact companion is proven without the need for any additional assumptions. This is possible, if such a binary is seen at high inclination. 
The project Massive Unseen Companions to Hot Faint Underluminous Stars from SDSS\footnote{Sloan Digital Sky Survey} (MUCHFUSS) aims at finding sdBs with compact companions like supermassive white dwarfs ($M>1.0\,{\rm M_{\odot}}$), neutron stars or black holes. First results of our follow-up campaign are published in Geier et al. (\cite{geier11}).

There is an interesting spin-off from this project: The same selection criteria that we applied to find such binaries are also well suited to single out hot subdwarf stars with constant high radial velocities in the Galactic halo like extreme population II stars or even hypervelocity stars. To refer to this aspect we coin the term Hyper-MUCHFUSS for the extended project. First results are presented in Tillich et al. (\cite{tillich11}). 

\section{Target selection}

The high fraction of sdB stars in close binary systems was initially discovered by the detection of RV shifts using time resolved spectroscopy (Maxted et al. \cite{maxted01}). In the past decade about 80 of these systems have been reobserved and their orbital parameters determined.  
We summarize the orbital parameters of all known sdB binaries and give references in Table~\ref{tab:orbitslit} (see also Fig.~\ref{periodK_lit}).

As far as the companion masses of the known sdB binaries could be constrained, it turned out that most companions should be either late main sequence stars with masses lower than half a solar mass or compact objects like white dwarfs. Targets for spectroscopic follow-up were selected in different ways dependent on the specific aims of each project.

For the MUCHFUSS project the target selection is optimised to find massive compact companions in close orbits around sdB stars. In order to discover rare objects applying the selection criteria explained in the forthcoming sections, a huge initial dataset is necessary. The enormous SDSS database (Data Release 6, DR6) is therefore the starting point for our survey. Best sky coverage is reached in the Northern hemisphere close to the galactic poles. SDSS data is widely used and therefore also well evaluated in terms of errors and accuracy (York et al. \cite{york00}; Abazajian et al. \cite{abazajian09}). The SDSS data are supplemented by additional spectroscopic observations of appropriate quality from other sources. 

\begin{table}[t!]
\caption{Survey observations. The first column lists the date of observation, while in the second
        the used telescope and instrumentation is shown. In the third column the initials of the observers are given.}
\label{survey-runs}
\begin{center}
\begin{tabular}{llll} \hline
\noalign{\smallskip}
Date & Telescope\,\&\,Instrument & Observers\\ \hline
\noalign{\smallskip}
January-June 2008 & CAHA-3.5m/TWIN & Service \\
2008/04/29--2008/05/01 & ING-WHT/ISIS & P. M., S. G., \\ 
& & S. B. \\ 
2008/08/13--2008/08/17 & CAHA-3.5m/TWIN & H. H. \\
2008/10/15--2008/10/19 & ESO-NTT/EFOSC2 & A. T. \\ 
April-July 2008 & ESO-VLT/FORS1 & Service \\ \hline
\end{tabular}
\end{center}
\end{table}

\begin{figure}[t!]
\begin{center}
	\resizebox{8cm}{!}{\includegraphics{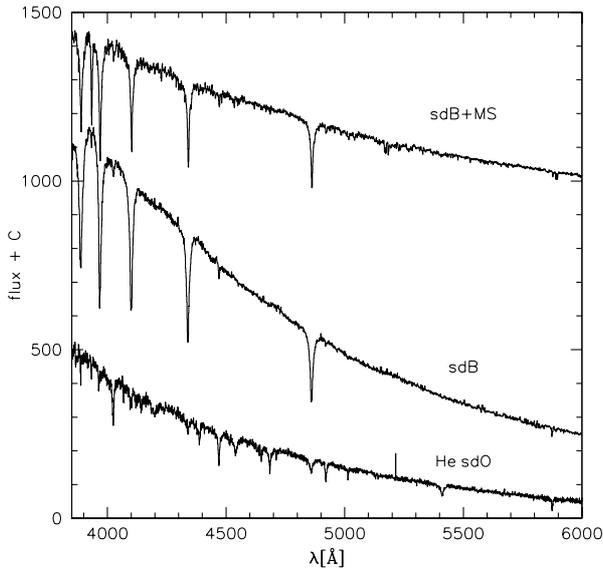}}
\end{center}
\caption{Flux calibrated SDSS spectra of a single-lined sdB, a helium rich sdO and an sdB with main sequence companion visible in the spectrum. Note the different slopes of the sdB and the sdB+MS spectra.}
\label{specexample}
\end{figure}

\subsection{Colour selection and visual classification}

Hot subdwarfs are found most easily by applying a colour cut to Sloan photometry. All spectra of point sources with colours $u-g<0.4$ and $g-r<0.1$ were selected. This colour criterion corresponds to a limit in the Johnson photometric system of $U-B<-0.57$ (Jester et al. \cite{jester05}), similar to the cut-off chosen by UV excess surveys, such as the Palomar Green survey (Green et al. \cite{green86}). The corresponding effective temperature of a BHB star is $\simeq15000\,{\rm K}$ (Castelli \& Kurucz \cite{castelli03}), well below the observed range for sdB stars ($>20000\,{\rm K}$). The limit of $g-r=+0.1$ corresponds to $B-V=+0.3$ (Jester et al. \cite{jester05}).
This ensures that sdBs in spectroscopic binaries are included if the dwarf companion is of spectral F or later, e.g. the sdB+F system PB\,8783 at $B-V=+0.13$ and $U-B=-0.65$ (Koen et al. \cite{koen97}). On the other hand the colour criteria exclude the huge number of QSOs (quasi stellar objects) which were the priority objects of SDSS in the first place. We selected $48\,267$ point sources with spectra in this way.

The spectra from SDSS are flux calibrated and cover the wavelength range from $3800\,{\rm \AA}$ to $9200\,{\rm \AA}$ with a resolution of $R=1800$. Rebassa-Mansergas et al. (\cite{rebassa07}) verified the wavelength stability to be $<14.5\,{\rm km\,s^{-1}}$ from repeat sub-spectra using SDSS observations of F-stars. We obtained the spectra of our targets from the SDSS Data Archive Server\footnote{das.sdss.org} and converted the wavelength scale from vacuum to air. The spectra were classified by visual inspection.

In a first step extragalactic objects, spectra with low quality ($S/N<5$) and unknown features have been excluded by visual inspection. In total we selected $10\,811$ spectra of $10\,153$ stars in this way. Fig.~\ref{colours_mag} (left panel) shows a two-colour plot of all selected objects. Classification was done by visual inspection of the spectra against reference spectra of hot subdwarfs and white dwarfs. Existence, width, and depth of helium and hydrogen absorption lines as well as the flux distribution between $4000$ and $6000\,{\rm \AA}$ were used as criteria. Subdwarf B stars show broadened hydrogen Balmer and He\,{\sc i} lines, sdOB stars He\,{\sc ii} lines in addition, while the spectra of sdO stars are dominated by weak Balmer and strong He\,{\sc ii} lines depending on the He abundance. A flux excess in the red compared to the reference spectrum as well as the presence of spectral features such as the Mg\,{\sc i} triplet at $5170\,{\rm \AA}$ or the Ca\,{\sc ii} triplet at $8650\,{\rm \AA}$ were taken as indications of a late type companion (for a few examples see Fig.~\ref{specexample}, for spectral classification of hot subdwarf stars see the review by Heber \cite{heber09}).

The sample contains $1100$ hot subdwarfs in total. $725$ belong to the class of single-lined sdBs and sdOBs. Because distinguising between these two subtypes from the spectral appearance alone can be difficult, we decided to treat them as one class. In addition we found $89$ sdBs with cool companions. $198$ stars are identified as single-lined sdOs, most of them enriched in helium. $9$ sdOs have a main sequence companion. In $79$ cases a unique classification was not possible. Most of these stars are considered as candidate sdBs of low temperature, which cannot be clearly distinguished from blue horizontal branch (BHB) stars or low mass white dwarfs of DA or DB type. 

\begin{figure}[t!]
\begin{center}
	\resizebox{8.5cm}{!}{\includegraphics{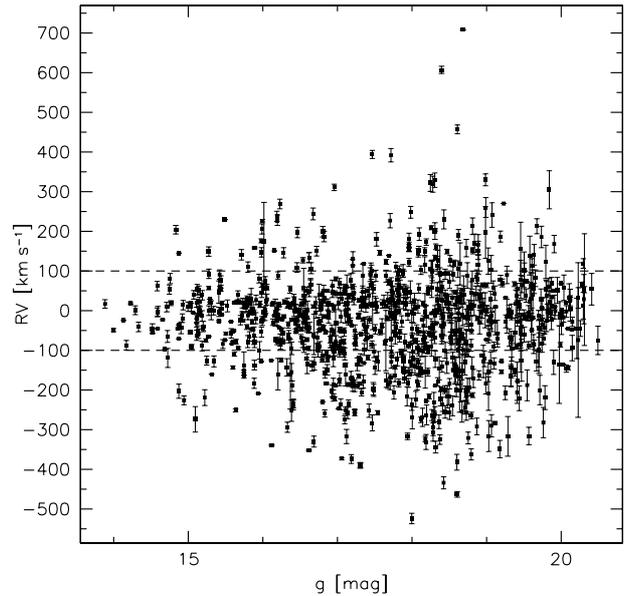}}
\end{center}
\caption{Heliocentric radial velocities of $1002$ subdwarfs plotted against $g$-magnitude. The two dashed lines mark the RV cut of $\pm100\,{\rm km\,s^{-1}}$.}
\label{RVall}
\end{figure}

Eisenstein et al. (\cite{eisenstein06}) used a semi-automatic method for the spectral classification of white dwarfs and hot subdwarfs from the SDSS DR4. It is instructive to compare their sample to ours. Our colour cut-off is more restrictive and the confusion limit ($S/N>5$) is brighter than that of Eisenstein et al. (\cite{eisenstein06}). Due to the redder colour cuts, blue horizontal branch stars enter the Eisenstein et al. sample, which we do not consider as hot subdwarf stars (see Heber \cite{heber09}). Applying our colour cuts to the hot subdwarf sample of Eisenstein et al. (\cite{eisenstein06}) yields 691 objects. The stars missing in our sample are mostly fainter than $g=19\,{\rm mag}$ as expected. Most recently, Kleinman (\cite{kleinman10}) extended the classifications to the SDSS DR7 and found 1409 hot subdwarf stars. Since no details are published, the sample can not be compared to ours yet. Considering our more restrictive colour cuts and confusion limit, the numbers compare very well with ours. This gives us confidence that our selection method is efficient.

In Fig.~\ref{colours_mag} (right panel) only the subdwarf stars brighter than $g=18\,{\rm mag}$ are plotted. With less pollution by poor spectra, two sequences become clearly visible. The solid symbols mark single-lined sdBs and sdOs, while the open squares mark binaries with late type companions of most likely K and G type visible in the spectra. The contribution of the cool companions shifts the colour of the star to the red. As can be seen in Figs.~\ref{colours_mag} the upper sequence also contains apparently single stars. Since the spectra are not corrected for interstellar reddening, some of these objects may show an excess in the red because of reddening rather than due to a cool companion. Amongst the faintest targets with noisy data, spectral features indicative of a late-type companion may have been missed as well as small excesses in the red.

In Fig.~\ref{colours_mag} (right panel) we also compare the sample to synthetic colours suitable for hot subdwarf stars. We chose the grid of Castelli \& Kurucz (\cite{castelli03})\footnote{http://wwwuser.oat.ts.astro.it/castelli/colors/sloan.html} and selected models with high gravity ($\log{g}=5.0$). The models reproduce the lower envelope of the targets in the colour-colour-diagram very well for effective temperatures ranging from $20\,000$ to $50\,000\,{\rm K}$ as expected for hot subdwarf stars. Different surface gravities, chemical compositions and interstellar reddening are not accounted for, but would explain the observed scatter of the stars.

It is interesting to note that there is an obvious lack of blue horizontal branch (BHB) stars with effective temperatures below $20\,000\,{\rm K}$ compared to the sdBs with higher temperatures. This gap is hardly caused by selection effects, because the BHB stars are brighter than the sdBs in the optical. We conclude that the number density of BHB stars in the analysed temperature range must be much smaller than the one of sdBs. Newell (\cite{newell73}) was the first to report the existence of such a gap in the two-colour diagram of field blue halo stars, which was subsequently found to be also present in some globular clusters (Momany et al. \cite{momany04}). The reason for this gap remains unclear (see the review by Catelan \cite{catelan09}). 

\begin{figure}[t!]
\begin{center}
	\resizebox{9.5cm}{!}{\includegraphics{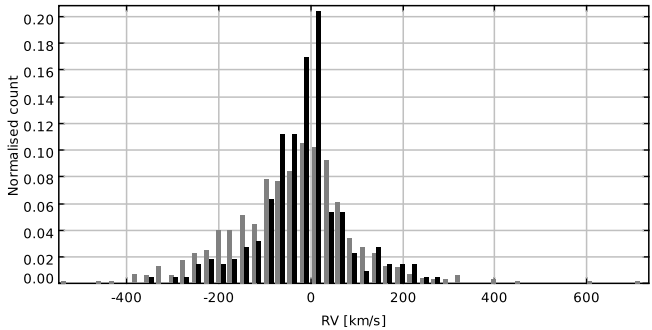}}
\end{center}
\caption{Radial velocity distribution of the hot subdwarf stars (see Fig.~\ref{RVall}). The bright sample ($g<16.5\,{\rm mag}$, black histogram) contains a mixture of stars from the disk and the halo population. The faint sample ($g>16.5\,{\rm mag}$, grey histogram) contains the halo population. The peak in the bright subsample around zero RV is caused by the thin disk population. The asymmetry in the faint subsample where negative RVs are more numerous than positive ones may be due to the presence of large structures in the halo and the movement of the solar system relative to the halo.}
\label{RVdistrib}
\end{figure}

\begin{figure}[t!]
\begin{center}
	\resizebox{8.5cm}{!}{\includegraphics{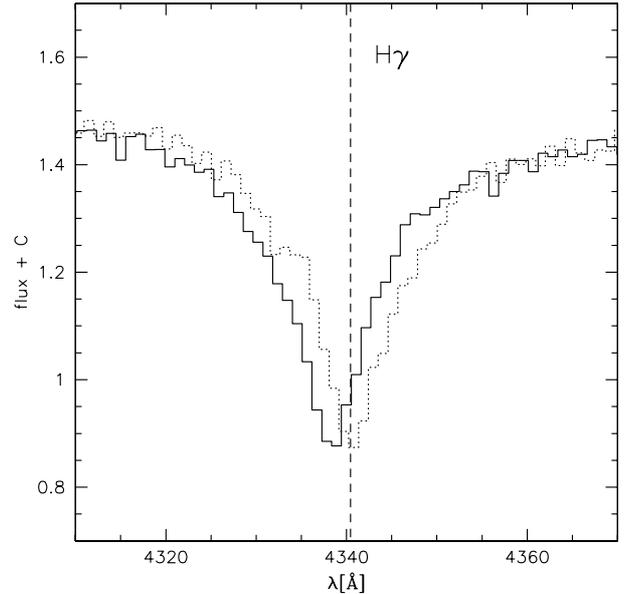}}
\end{center}
\caption{H$_{\rm \beta}$-line of two consecutively taken individual SDSS spectra ($\Delta t=0.056\,{\rm d}$) of the sdB binary J113840.68$-$003531.7. The shift in RV ($\simeq140\,{\rm km\,s^{-1}}$) between the two exposures is clearly visible.}
\label{specvar}
\end{figure}

\subsection{High radial velocity sample (HRV)}

The radial velocities of all identified hot subdwarf stars {both single- and double-lined} were measured by fitting a set of mathematical functions (Gaussians, Lorentzians and polynomials) to the hydrogen Balmer lines as well as helium lines if present using the FITSB2 routine (Napiwotzki et al. \cite{napiwotzki04a}) and the Spectrum Plotting and Analysis Suite (SPAS) developed by H. Hirsch. Fig.~\ref{RVall} shows the RVs of $1002$ hot subdwarf stars. 

Most of the known sdB binaries are bright objects ($V\simeq10-14\,{\rm mag}$) and the vast majority of them belongs to the Galactic disk population (Altmann et al. \cite{altmann04}). Due to the fact that these binary systems are close to the Sun they rotate around the Galactic centre with approximately the same velocity. That is why the system velocities of most sdB binaries relative to the Sun are low. One quarter of the known systems have $|\gamma|<10\,{\rm km\,s^{-1}}$, $85\%$ have $|\gamma|<50\,{\rm km\,s^{-1}}$ (see Table~\ref{tab:orbitslit}). Furthermore the RV semiamplitudes of these binaries are in most cases lower than $100\,{\rm km\,s^{-1}}$ (see Fig.~\ref{periodK_lit}). In order to filter out such normal thin-disk binaries we excluded sdBs with RVs lower than $\pm100\,{\rm km\,s^{-1}}$. 

Typical hot subdwarf stars fainter than $g\simeq17\,{\rm mag}$ have distances exceeding $4\,{\rm kpc}$ and therefore likely belong to the Galactic halo population. Most of the stars in our sample are fainter than that (see Fig.~\ref{RVall}). The velocity distribution in the halo is roughly consistent with a Gaussian of $120\,{\rm km\,s^{-1}}$ dispersion (Brown et al. \cite{brown05}). Fig.~\ref{RVdistrib} shows the velocity distribution of our sample dependent on the brightness of the objects. The distribution of the bright subsample ($g<16.5\,{\rm mag}$) is roughly similar to the one of the faint subsample ($g>16.5\,{\rm mag}$), the later extending to more extreme velocities and being somewhat asymmetric. Selecting objects with heliocentric radial velocities exceeding $\pm100\,{\rm km\,s^{-1}}$ we aim at finding halo stars with extreme kinematics as well as close binaries with high RV amplitudes.

Another selection criterion is the brightness of the stars. The accuracy of the RV measurements depends on the S/N of the spectra and the existence and strength of the spectral lines. Furthermore, the classification becomes more and more uncertain as soon as the S/N drops below $\simeq10$ and the probability of including DAs rises. Objects of uncertain type and RV (errors larger than $50\,{\rm km\,s^{-1}}$) have therefore been excluded. Most of the excluded objects are fainter than $g=19\,{\rm mag}$. Altogether the target sample consists of $258$ stars. 

Second epoch medium resolution spectroscopy was obtained starting in 2008 using ESO-VLT/FORS1 ($R\simeq1800,\lambda=3730-5200\,{\rm \AA}$), WHT/ISIS  ($R\simeq4000,\lambda=3440-5270\,{\rm \AA}$), CAHA-3.5m/TWIN ($R\simeq4000,\lambda=3460-5630\,{\rm \AA}$) and ESO-NTT/EFOSC2 ($R\simeq2200,\lambda=4450-5110\,{\rm \AA}$). The journal of observations is given in Table~\ref{survey-runs}. Up to now we have reobserved $88$ stars. We discovered $\simeq30$ halo star candidates with constant high radial velocity (see Tillich et al. \cite{tillich11}) as well as $46$ systems with radial velocities that were most likely variable.

\begin{figure}[t!]
\begin{center}
	\resizebox{8.5cm}{!}{\includegraphics{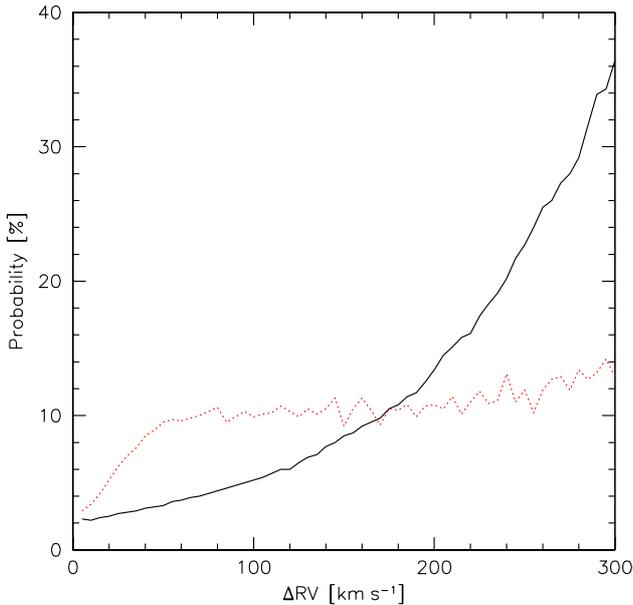}}
\end{center}
\caption{Probability for an sdB binary to host a massive compact companion and to be seen at sufficiently high inclination to unambiguously identify it from its binary mass function plotted against the RV shift within random times (solid curves, HRV sample) or on short timescales (dotted curve, RRV sample).}
\label{muchfuss_sim}
\end{figure}

\subsection{Rapid radial velocity variable sample (RRV)}

All SDSS spectra are co-added from at least three individual ``sub-spectra'' with typical exposure times of 15\,min. In most
  cases, the sub-spectra are taken consecutively; however, occasionally they may be split over several nights. In addition,
  some SDSS objects are observed more than once, either because the
  entire spectroscopic plate is re-observed, or because they are in
  the overlap area between adjacent spectroscopic plates. As a result, up
  to 30 sub-spectra are available for some objects. Hence, SDSS
  spectroscopy can be used to probe for radial velocity variations, a
  method pioneered by Rebassa-Mansergas et al. (\cite{rebassa07}) to identify
  close white dwarf plus main-sequence binaries. We have obtained the sub-spectra for
  all sdBs brighter than $g=18.5\,{\rm mag}$ from the SDSS Data Archive
  Server. The quality of the individual spectra is not sufficient for our analysis in the case of even fainter stars. 
The object spectra were extracted from the FITS files for the blue and red spectrographs, and merged
  into a single spectrum using \texttt{MIDAS}. From the inspection of
  these data, we discovered $81$ new candidate sdB binaries with radial
  velocity variations on short time scales, $\simeq0.02-0.07\,{\rm d}$ (see
  Fig.\,~\ref{specvar} for an example). 

The individual SDSS spectra are perfectly suited to search for close double degenerate binaries. Ongoing projects like SWARMS (Badenes et al. \cite{badenes09}; Mullally et al. \cite{mullally09}) focus on binaries with white dwarf primaries (see also Kilic et al. \cite{kilic10}; Marsh et al. \cite{marsh10}) and use a similar method. 

\begin{figure}[t!]
\begin{center}
	\resizebox{8.5cm}{!}{\includegraphics{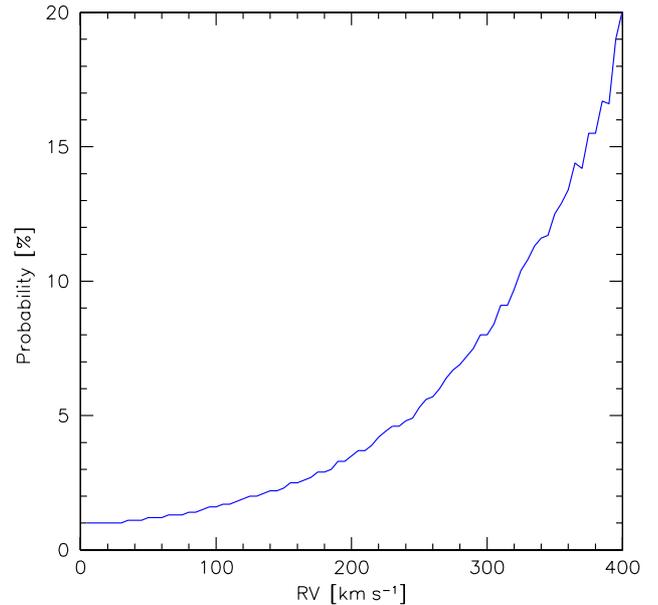}}
\end{center}
\caption{Same as Fig.~\ref{muchfuss_sim} except that the probability is plotted against RV at random time.}
\label{muchfuss_sim_const}
\end{figure}

\subsection{Selecting high mass companions}

Time resolved follow-up spectroscopy with a good phase coverage is needed to determine the orbital solutions of the RV variable systems. In order to select the most promising targets for follow-up, we carried out numerical simulations and estimated the probability for a subdwarf binary with known RV shift to host a massive compact companion. We created a mock sample of sdBs with a close binary fraction of $50\,\%$. 

We adopted the distribution of orbital periods of all known sdB binaries (see Table~\ref{tab:orbitslit}) approximated by two Gaussians centered at $0.7\,{\rm d}$ (width $0.3\,{\rm d}$) and $5.0\,{\rm d}$ (width $3.0\,{\rm d}$) days and assumed that $82\%$ of the binaries belong to the short period population. The short period Gaussian was truncated at $0.05\,{\rm d}$, which is considered the minimum period for an sdB binary, because the subdwarf primary starts filling its Roche lobe for shorter periods and typical companion masses. Since stable Roche lobe overflow and the accretion onto the companion would dramatically change the spectra of these stars,
we can safely presume that our sample does not contain such objects.

The orbital inclination angles are assumed to be randomly distributed, but for geometrical reasons binaries at high inclinations are more likely observed than binaries at low inclinations. To account for this, we used the method described in Gray (\cite{gray92}) and adopted a realistic distribution of inclination angles. 

For the sdB mass the canonical value of $0.47\,{\rm M_{\odot}}$ was chosen. The distribution of companion masses was based on the results by Geier et al. (\cite{geier10b}). The distribution of the low mass companions was approximated by a Gaussian centered at $0.4\,{\rm M_{\odot}}$ (width $0.3\,{\rm M_{\odot}}$). The fraction of massive compact companions is estimated to $2\,\%$ of the close binary population based on binary population synthesis models (Geier et al. \cite{geier10b}). The mass distribution of these companions was approximated by a Gaussian centered at $2.0\,{\rm M_{\odot}}$ (width $1.0\,{\rm M_{\odot}}$). 

For the system velocities a Gaussian distribution with a dispersion of $120\,{\rm km\,s^{-1}}$ typical for halo stars was adopted (Brown et al. \cite{brown05}). Two RVs were taken from the model RV curves at random times and the RV difference was calculated for each of the $10^{6}$ binaries in the simulation sample. This selection criterion corresponds to the HRV sample. For given RV difference and timespan between the measurements the fraction of systems with minimum companion masses exceeding $1\,{\rm M_{\odot}}$ was counted.

In Fig.~\ref{muchfuss_sim} the fraction of massive compact companions with unambiguous mass functions is plotted against the RV shift between two measurements at random times (solid curve). It is quite obvious that binaries with high RV shifts are more likely to host massive companions. The probability for a high mass companion ($>1\,{\rm M_{\odot}}$) at high inclination is raised by a factor of ten as soon as the RV shift exceeds $200\,{\rm km\,s^{-1}}$. 

In order to check whether the selection of high velocities rather than high velocity shifts has an impact on the probability of finding sdB binaries with massive compact companions we used the same simulation. In Fig.~\ref{muchfuss_sim_const} the fraction of these binaries is plotted against only one RV measurement taken at a random time. It can be clearly seen that the detection probability rises significantly for stars with high RVs. Selecting the fastest stars in the halo therefore makes sense when searching for massive compact companions to sdB.

Since the individual SDSS spectra were taken within short timespans, another simulation was performed corresponding to the RRV sample. The first RV was taken at a random time, but the second one just $0.03\,{\rm d}$ later. The dotted curve in Fig.~\ref{muchfuss_sim} illustrates the outcome of this simulation. As soon as the RV shift exceeds $30\,{\rm km\,s^{-1}}$ within $0.03\,{\rm d}$ the probability that the companion is massive rises to $\simeq10\%$. The reason why the probability does not increase significantly with increasing RV shift is that the most massive companions in our simulation have maximum RV shifts as high as $1000\,{\rm km\,s^{-1}}$. At the most likely periods of $\simeq0.5\,{\rm d}$ the maximum RV shift within $0.03\,{\rm d}$ is then of the order of $100\,{\rm km\,s^{-1}}$. Even higher RV shifts within short time are not physically plausible.

\begin{figure}[t!]
\begin{center}
	\resizebox{8cm}{!}{\includegraphics{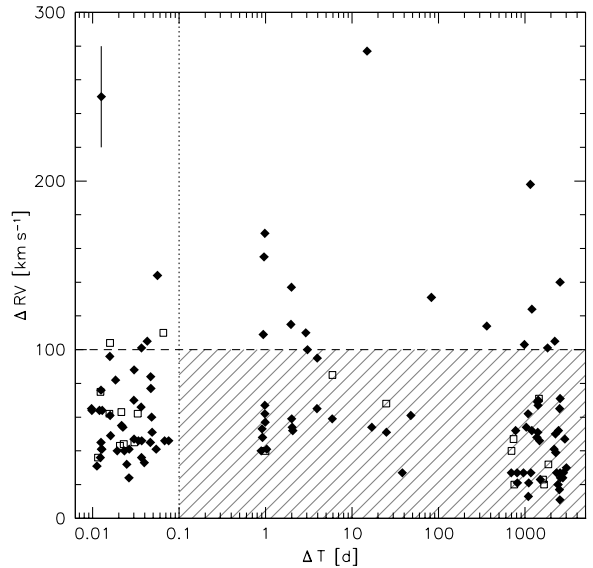}}
\end{center}
\caption{Highest radial velocity shift between individual spectra plotted against time difference between the corresponding observing epochs. The dashed horizontal line marks the selection criterion $\Delta RV>100\,{\rm km\,s^{-1}}$, the dotted vertical line the selection criterion $\Delta T<0.1\,{\rm d}$. All objects fulfilling at least one of these criteria lie outside the shaded area and belong to the top candidate list for the follow-up campaign. The filled diamonds mark sdBs, while the blank squares mark He-sdOs.}
\label{dTdRV}
\end{figure}

Our simulation gives a quantitative estimate based on our current knowledge of the sdB binary populations. It has to be pointed out that these numbers should be considered as rough estimates at most. The observed period and companion mass distributions are especially  affected by selection effects. The derived numbers are therefore only used to create a priority list and select the best targets for follow-up.

\subsection{Final target sample}

Our sample of promising targets consists of $69$ objects in total. $52$ stars show significant RV shifts ($>30\,{\rm km\,s^{-1}}$) within $0.02-0.07\,{\rm d}$ and are selected from the RRV sample, while $17$ stars show high RV shifts ($100-300\,{\rm km\,s^{-1}}$) within more than one day and are selected from the HRV sample (see Fig.~\ref{dTdRV}).

In Geier et al. (\cite{geier11}) we showed that the SDSS spectra are well suited to determine atmospheric parameters by fitting synthetic line profiles to the hydrogen Balmer lines (H$_{\rm \beta}$ to H$_{\rm 9}$) as well as He\,{\sc i} and He\,{\sc ii} lines. In order to maximize the quality of the data the single spectra were shifted to rest wavelength and coadded. The quality of the averaged spectra is quite inhomogeneous ($S/N\simeq20-180$, see Table\,\ref{targets:quality}), which affects the accuracy of the parameter determination.

\begin{figure*}[t!]
\begin{center}
	\resizebox{6cm}{!}{\includegraphics{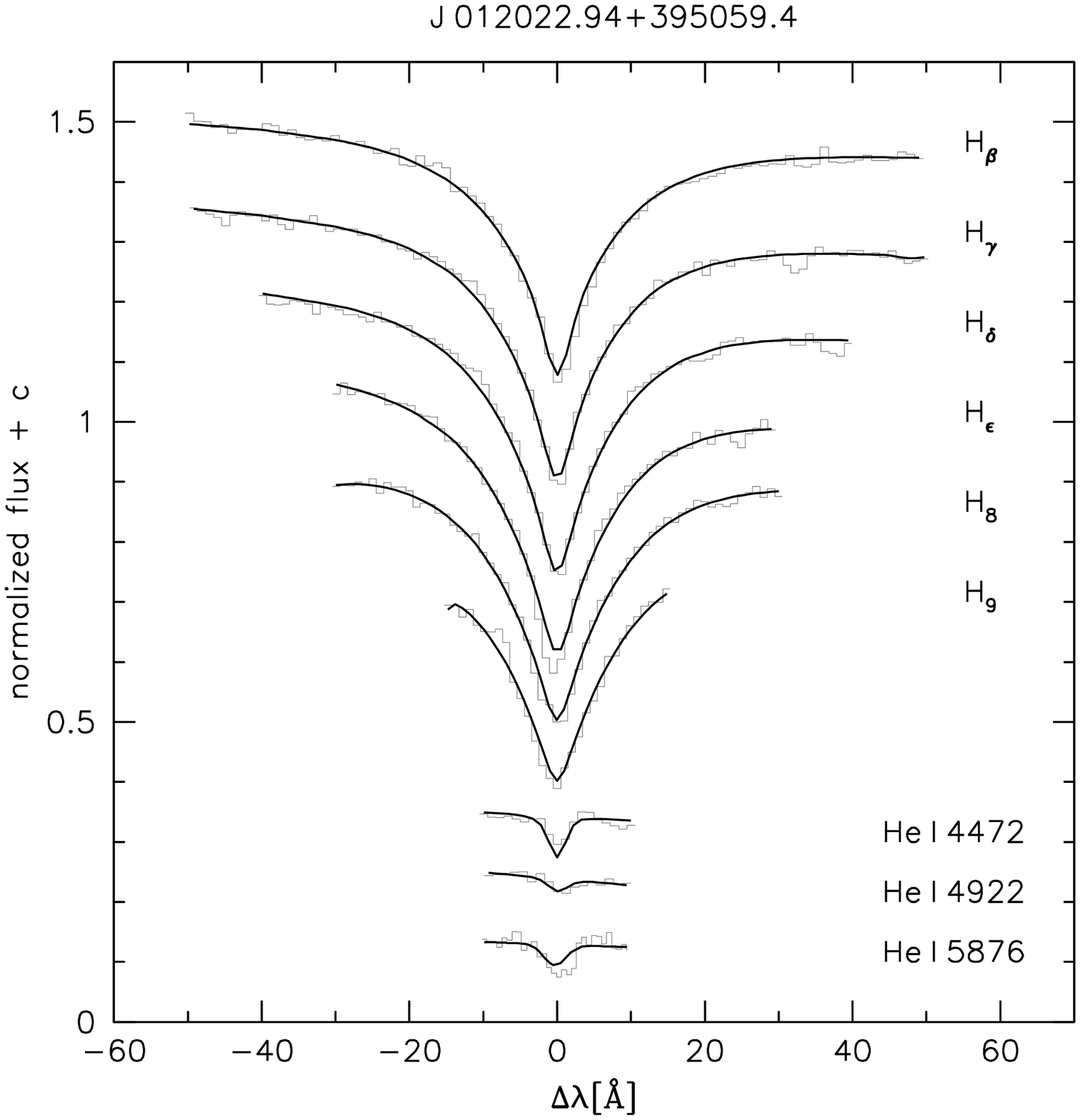}}
	\resizebox{6cm}{!}{\includegraphics{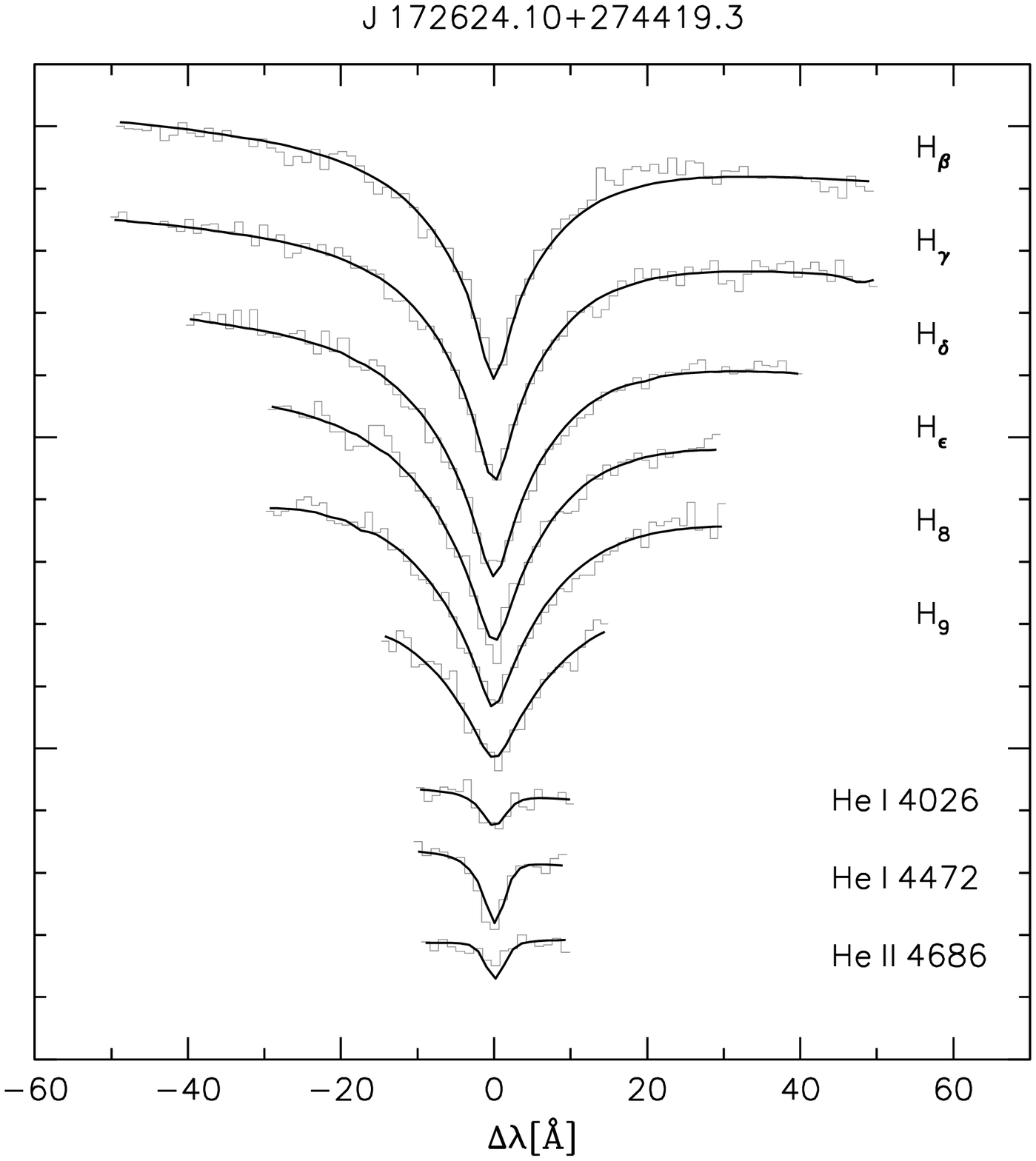}}
	\resizebox{6cm}{!}{\includegraphics{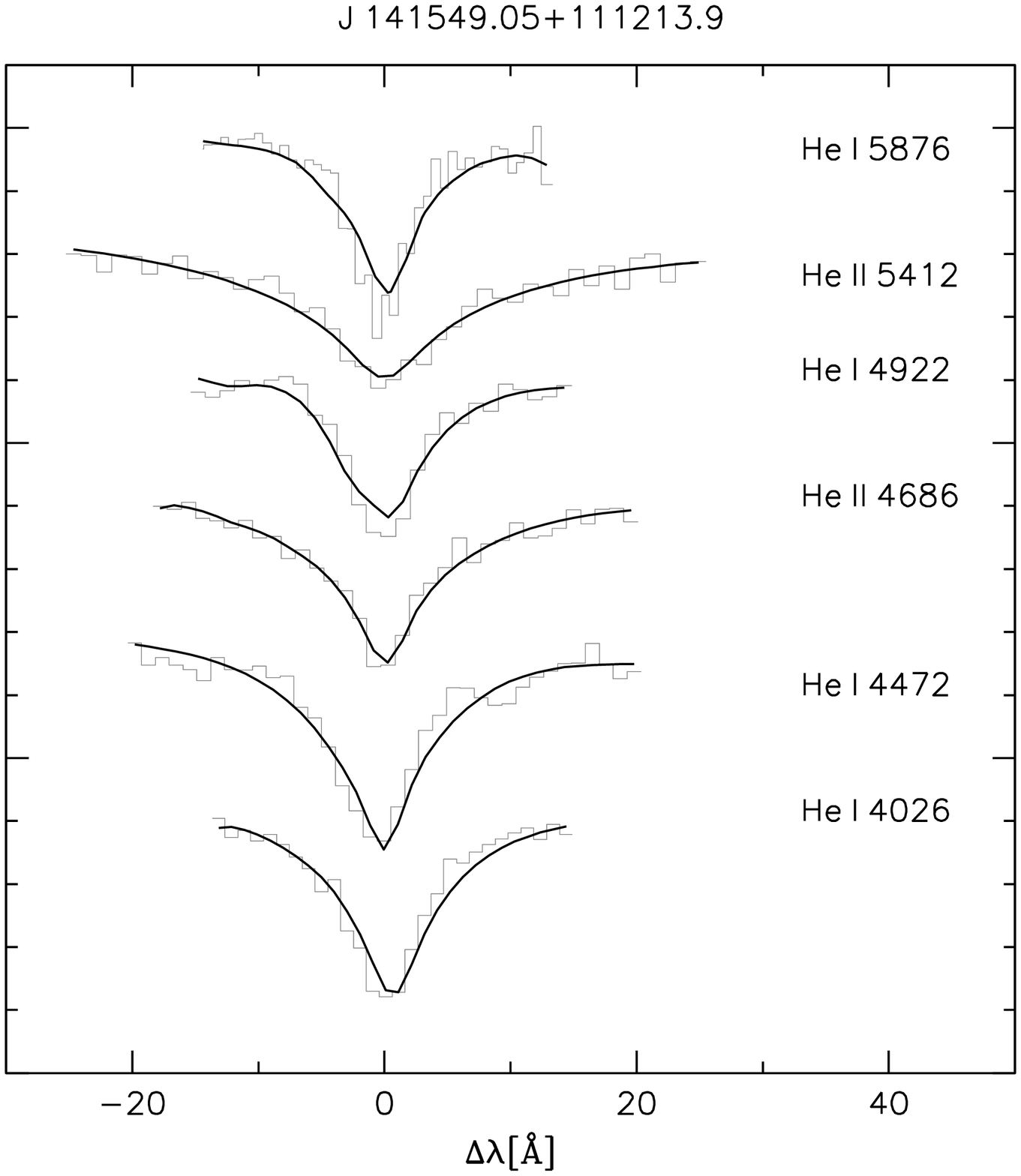}}
\end{center}
\caption{Example fits of hydrogen and helium lines with model spectra for an sdB (left panel), an sdOB (middle panel) and a He-sdO star (right panel). The atmospheric parameters of these stars are given in Tables~\ref{targets:hrv} and \ref{targets:rrv}.}
\label{fit_muchfuss}
\end{figure*}

A quantitative spectral analysis was performed in the way described in Lisker et al. (\cite{lisker05}) and Str\"oer et al. (\cite{stroeer07}). Due to the fact that our sample consists of different subdwarf classes, we used appropriate model grids in each case. For the  hydrogen-rich and helium-poor ($\log{y}<-1.0$) sdBs with effective temperatures below $30\,000\,{\rm K}$ a grid of metal line blanketed LTE atmospheres with solar metallicity was used. Helium-poor sdBs and sdOBs with temperatures ranging from $30\,000\,{\rm K}$ to $40\,000\,{\rm K}$ have been analysed using LTE models with enhanced metal line blanketing (O'Toole \& Heber \cite{otoole06}). In the case of hydrogen-rich sdOBs with temperatures below $40\,000\,{\rm K}$ showing moderate He-enrichment ($\log{y}=-1.0..0.0$) and hydrogen-rich sdOs metal-free NLTE models were used (Str\"oer et al. \cite{stroeer07}). The He-sdOs have been analysed with NLTE models taking into account the line-blanketing caused by nitrogen and carbon (Hirsch \& Heber \cite{hirsch09}).

Spectral lines of hydrogen and helium were fitted by means of $\chi^{2}$ minimization using SPAS. The statistical errors have been calculated with a bootstrapping algorithm. Minimum errors reflecting systematic shifts when using different model grids ($\Delta T_{\rm eff}=500\,{\rm K}$; $\Delta \log{g}=0.05$; $\Delta \log{y}=0.1$, for a discussion see Geier et al. \cite{geier07}) have been adopted in cases where the statistical errors were lower. Example fits for a typical sdB, an sdOB and a He-sdO star are shown in Fig.~\ref{fit_muchfuss}.

In addition to statistical uncertainities systematic effects have to be taken into account in particular for sdB stars. The higher Balmer lines (H$_{\rm \epsilon}$ and higher) at the blue end of the spectral range are very sensitive to changes in the atmospheric parameters. However, the SDSS spectral range restricts our analysis to the Balmer lines from H$_{\rm \beta}$ to H$_{\rm 9}$. In high S/N data these lines are sufficient to measure accurate parameters as has been shown in Geier et al. (\cite{geier11}). In spectra of lower quality the bluest lines (H$_{\rm 9}$ and H$_{\rm 8}$) are dominated by noise and cannot be used any more. In order to check whether this leads to systematic shifts in the parameters as reported in Geier et al. (\cite{geier10b}) we made use of the individual SDSS spectra. We chose objects with multiple spectra, which have an S/N comparable to the lowest quality data in our sample ($\simeq20$). The atmospheric parameters were obtained from each individual spectrum. Average values of $T_{\rm eff}$ and $\log{g}$ were calculated and compared to the atmospheric parameters derived from the analysis of the appropriate coadded spectrum. For effective temperatures ranging from $27\,000\,{\rm K}$ and $39\,000\,{\rm K}$ no significant systematic shifts were found. This means that the error is dominated by statistical noise. However, for temperatures as low as $25\,000\,{\rm K}$ systematic shifts of the order of $-2500\,{\rm K}$ in $T_{\rm eff}$ and $-0.35$ in $\log{g}$ are present. For sdBs with low effective temperatures and signal-to-noise, the atmospheric parameters are therefore systematically underestimated. Only three stars in our sample have temperatures in this range. Since their coadded spectra are of reasonable quality ($S/N=34-167$), systematic shifts should be negligible in these cases. Because all important lines of He\,{\sc i} and He\,{\sc ii} are well covered by the SDSS spectral range, systematic effects should be negligible in the case of He-rich sdO/Bs as well.

The parameters of the sample are given in Tables\,\ref{targets:hrv} and \ref{targets:rrv}. Seven stars have already been analysed in Geier et al. (\cite{geier11}). The sample consists of $38$ hydrogen rich sdBs, $13$ sdOBs and $3$ hydrogen rich sdOs. Thirteen stars are helium rich sdOs (He-sdOs) and J134352.14+394008.3 belongs to the rare class of helium rich sdBs.

Our SDSS sample reaches down to fainter magnitudes and hence, larger distances than any previous survey. In an ongoing project Green et al. (\cite{green08}) analyse all hot subdwarfs from the PG survey down to $\simeq14.0\,{\rm mag}$. The sample of hot subdwarf stars analysed in the course of the SPY survey reaches down to $\simeq16.5\,{\rm mag}$ (Lisker et al. \cite{lisker05}; Str\"oer et al. \cite{stroeer07}), quite similar to the sample of sdBs from the Hamburg Quasar Survey analysed by Edelmann et al. (\cite{edelmann03}). 

Spectroscopic distances to our stars have been calculated as described in Ramspeck et al. (\cite{ramspeck01}) assuming the canonical mass of $0.47\,{\rm M_{\odot}}$ for the subdwarfs and using the formula given by Lupton\footnote{http://www.sdss.org/dr6/algorithms/sdssUBVRITransform.html} to convert SDSS-g and r magnitudes to Johnson V magnitudes. Again interstellar reddening has been neglected. The distances range from $1\,{\rm kpc}$ to $>16\,{\rm kpc}$. Since the SDSS footprint is roughly perpendicular to the Galactic disk, these distances tell us something about the population membership of our stars. These subdwarfs most likely belong to the thick disk or the halo with small contributions of thin disk stars. 

\begin{figure}[t!]
\begin{center}
	\resizebox{8.5cm}{!}{\includegraphics{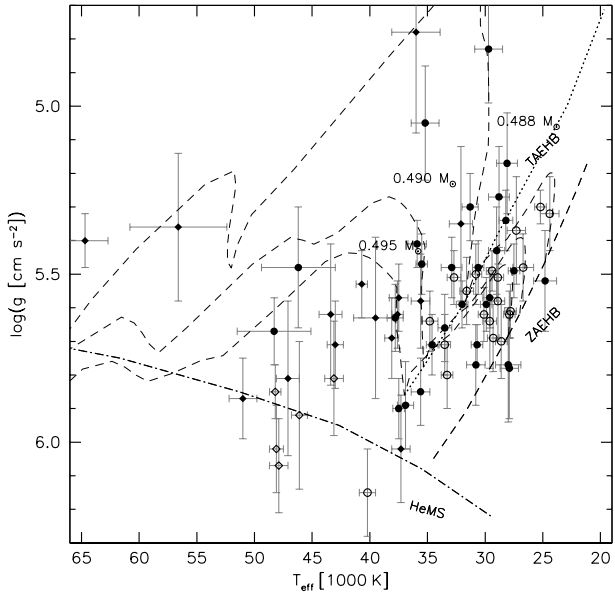}}
\end{center}
\caption{$T_{\rm eff}-\log{g}$ diagram of our target sample. The helium main sequence (HeMS) and the EHB band (limited by the zero-age EHB, ZAEHB, and the terminal-age EHB, TAEHB) are superimposed with EHB evolutionary tracks for subsolar metallicity ($\log{z}=-1.48$) from Dorman et al. (\cite{dorman93}).}
\label{tefflogg_mlow}
\end{figure}

\begin{figure}[t!]
\begin{center}
	\resizebox{8.5cm}{!}{\includegraphics{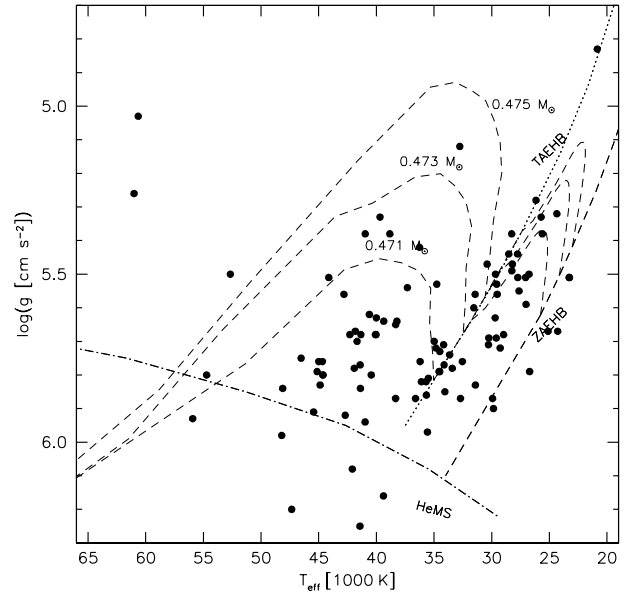}}
\end{center}
\caption{$T_{\rm eff}-\log{g}$ diagram of the hot subdwarfs from the SPY project (Lisker et al. \cite{lisker05}; Str\"oer et al. \cite{stroeer07}). The helium main sequence (HeMS) and the EHB band (limited by the zero-age EHB, ZAEHB, and the terminal-age EHB, TAEHB) are superimposed with EHB evolutionary tracks for solar metallicity from Dorman et al. (\cite{dorman93}).}
\label{tefflogg_spy}
\end{figure}

Fig.~\ref{tefflogg_mlow} shows a $T_{\rm eff}-\log{g}$ diagram of the top target sample. Most of our stars were born in an environment of low metallicity (thick disk or halo). Dorman et al. (\cite{dorman93}) calculated evolutionary tracks for different metallicities of the subdwarf progenitor stars. For lower metallicities, the evolutionary tracks and with them the location of the EHB, are shifted towards higher temperatures and lower surface gravities. In Fig.~\ref{tefflogg_mlow} the $T_{\rm eff}-\log{g}$ diagram is superimposed with evolutionary tracks and an EHB calculated for a subsolar metallicity of $\log{z}=-1.48$, which is consistent with a mixture between thick disk and halo population. Evolutionary tracks for solar metallicity are given in Fig.~\ref{tefflogg_spy} for comparison.

Most of the sdB stars with hydrogen-rich atmospheres are found on or slightly above the EHB band implying an evolutionary status as core helium-burning EHB or shell helium-burning post-EHB stars. The sample contains only three hydrogen rich sdOs, which are thought to be evolved post-EHB stars in a transition state. The He-sdOs cluster near the HeMS at temperatures of $\simeq45\,000\,{\rm K}$. This is fully consistent with the results from the PG and the SPY surveys (Green et al. \cite{green08}; Lisker et al. \cite{lisker05}; Str\"oer et al. \cite{stroeer07}) and illustrates that our sample is not biased (see Fig.~\ref{tefflogg_spy}). 

Compared to other studies we find only a few stars with temperatures lower than $27\,000\,{\rm K}$. Furthermore, the scatter around the EHB seems to be systematically shifted towards higher temperatures and lower surface gravities. According to our study of systematic errors in the parameter determination, it is unlikely that this causes the effect. However, higher quality data would be necessary to verify this. Another possible explanation might be related to the volume of the sample. Since hot subdwarfs of lower temperature are brighter in the optical range because of the lower bolometric correction, we may already see all of them in a fixed volume, while the fraction of hot stars is still rising at fainter magnitudes.

\begin{figure}[h!]
\begin{center}
	\resizebox{8.5cm}{!}{\includegraphics{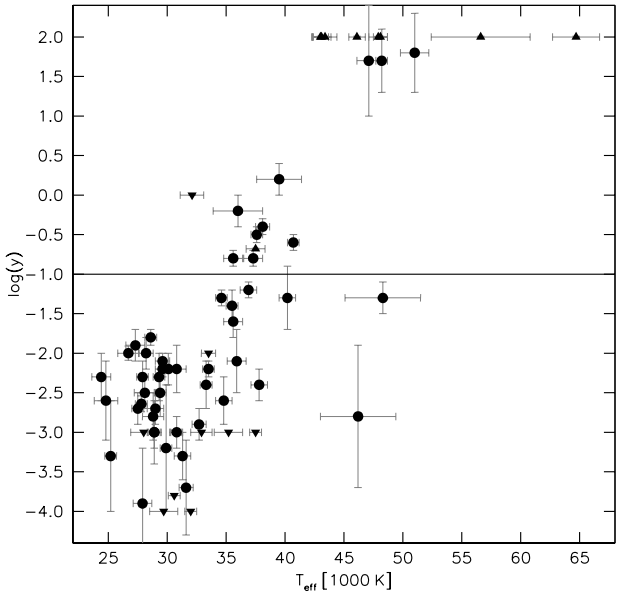}}
\end{center}
\caption{Helium abundance $\log{y}$ plotted against effective temperature (see Tables~\ref{targets:hrv},\ref{targets:rrv}). The solid horizontal line marks the solar value. Lower and upper limits are marked with upward and downward triangles.}
\label{nhevsteff}
\end{figure}

In Fig.~\ref{nhevsteff} the helium abundance is plotted against effective temperature. The general correlation of helium abundance with effective temperature and the large scatter in the region of the sdB stars have been observed in previous studies as well. Two sequences of helium abundance among the sdB stars as reported by Edelmann et al. (\cite{edelmann03}) could not be identified. 

One has to keep in mind that our sample consists of RV variable stars only. In Fig.~\ref{tefflogg_mlow} a lack of such stars at the hot end of the EHB is visible. Green et al. (\cite{green08}) reported similar systematics in their bright PG sample. The reason for this behaviour is not fully  understood yet. According to the model of Han et al. (\cite{han02,han03}) and Han (\cite{han08}) sdBs with thin hydrogen envelopes situated at the hot end of the EHB may be formed after the merger of two helium WDs. Since merger remnants are single stars, they are not RV variable. 

The top target sample includes 13 He-sdOs where RV shifts of up to $100\,{\rm km\,s^{-1}}$ have been detected within short timespans of $0.01-0.1\,{\rm d}$. In total 20 He-sdOs show signs of RV variability. This fraction was unexpected since the fraction of close binary He-sdOs from the SPY sample turned out to be $4\%$ at most (Napiwotzki \cite{napiwotzki08}).\footnote{Green et al. (\cite{green08}) suggested that the binary fraction of He-sdO stars may be comparable to the binary fraction of sdBs.} 

\section{Summary and Outlook}

In this paper we introduced the MUCHFUSS project, which aims at finding sdBs in close binaries with massive compact companions. We identified $1100$ hot subdwarf stars from the SDSS by colour selection and visual inspection of their spectra. Stars with high absolute radial velocities have been selected to efficiently remove normal sdB binaries from the thin disk population and have been reobserved. $46$ binary candidates with significant RV shifts have been found. From the analysis of individual SDSS spectra, $81$ additional stars with RV shifts on short timescales have been found.

Targets for follow-up spectroscopy have been selected using numerical simulations based on the properties of the known sdB close binary population and theoretical predictions about the relative fraction of massive compact companions. $69$ binaries with high RV shifts as well as significant RV shifts on short timescales have been selected as good candidates for massive compact companions. Atmospheric parameters, spectroscopic distances and population memberships have been determined. 

The multi-site follow-up campaign started in 2009 and is being conducted with medium resolution spectrographs mounted at several different telescopes of mostly 4-m-class. First results are presented in Geier et al. (\cite{geier11}).

\begin{acknowledgements}

A.T., S.G. and H.H. are supported by the Deutsche Forschungsgemeinschaft (DFG) through grants HE1356/45-1, HE1356/49-1, and HE1356/44-1, respectively. R.\O. acknowledges funding from the European Research Council under the European Community's Seventh Framework Programme (FP7/2007--2013)/ERC grant agreement N$^{\underline{\mathrm o}}$\,227224 ({\sc prosperity}), as well as from the Research Council of K.U.Leuven grant agreement GOA/2008/04.

Funding for the SDSS and SDSS-II has been provided by the Alfred P. Sloan Foundation, the Participating Institutions, the National Science Foundation, the U.S. Department of Energy, the National Aeronautics and Space Administration, the Japanese Monbukagakusho, the Max Planck Society, and the Higher Education Funding Council for England. The SDSS Web Site is http://www.sdss.org/.

The SDSS is managed by the Astrophysical Research Consortium for the Participating Institutions. The Participating Institutions are the American Museum of Natural History, Astrophysical Institute Potsdam, University of Basel, University of Cambridge, Case Western Reserve University, University of Chicago, Drexel University, Fermilab, the Institute for Advanced Study, the Japan Participation Group, Johns Hopkins University, the Joint Institute for Nuclear Astrophysics, the Kavli Institute for Particle Astrophysics and Cosmology, the Korean Scientist Group, the Chinese Academy of Sciences (LAMOST), Los Alamos National Laboratory, the Max-Planck-Institute for Astronomy (MPIA), the Max-Planck-Institute for Astrophysics (MPA), New Mexico State University, Ohio State University, University of Pittsburgh, University of Portsmouth, Princeton University, the United States Naval Observatory, and the University of Washington. 


\end{acknowledgements}

\begin{table*}[t!]
\caption{Priority targets for follow-up. Besides the names, the g magnitudes, the number of individual spectra and the S/N of the coadded spectra at $\simeq4100\,{\rm \AA}$ are given.}
\label{targets:quality}
\begin{center}
\begin{tabular}{llllllllll} \hline
\noalign{\smallskip}
Object            &        &  g     & No.  & S/N   &  Object            &        &  g     & No.  & S/N \\ 
                  &        &   &         &                        &        &   &         & \\ \hline
\noalign{\smallskip}
J002323.99$-$002953.2 & PB\,5916           & $15.3$ & 16 & 116 & J150513.52+110836.6   & PG\,1502+113       & $15.1$ & 4  & 90      \\   
J012022.94+395059.4   & FBS\,0117+396      & $15.2$ & 8  & 100 & J150829.02+494050.9   &                    & $17.3$ & 3  & 50      \\   
J012739.35+404357.8   &                    & $16.5$ & 8  & 59  & J151415.66$-$012925.2 &                    & $16.8$ & 5  & 48      \\   
J052544.93+630726.0   &                    & $17.6$ & 3  & 35  & J152222.15$-$013018.3 &                    & $17.7$ & 5  & 28       \\  
J074534.16+372718.5   &                    & $17.6$ & 5  & 26  & J152705.03+110843.9   &                    & $17.1$ & 5  & 39       \\  
J075937.15+541022.2   &                    & $17.5$ & 3  & 27  & J153411.10+543345.2   & WD\,1532+547       & $16.7$ & 8  & 52       \\  
J082053.53+000843.4   &                    & $14.9$ & 6  & 103 & J155628.34+011335.0   &                    & $16.0$ & 8  & 92       \\  
J083006.17+475150.4   &                    & $15.8$ & 5  & 95  & J161140.50+201857.0   &                    & $18.2$ & 5  & 20       \\  
J085727.65+424215.4   & US\,1993           & $18.3$ & 4  & 21  & J161817.65+120159.6   &                    & $17.8$ & 4  & 18       \\  
J092520.70+470330.6   &                    & $17.4$ & 3  & 33  & J162256.66+473051.1   &                    & $16.0$ & 4  & 72       \\  
J094856.95+334151.0   & KUV\,09460+3356    & $17.4$ & 3  & 46  & J163702.78$-$011351.7 &                    & $17.1$ & 12 & 46       \\  
J095229.62+301553.6   &                    & $18.2$ & 3  & 20  & J164326.04+330113.1   & PG\,1641+331       & $16.1$ & 3  & 55       \\  
J095238.93+252159.7   &                    & $14.5$ & 4  & 113 & J165404.26+303701.8   & PG\,1652+307       & $15.1$ & 4  & 167      \\  
J100535.76+223952.1   &                    & $18.1$ & 4  & 28  & J170645.57+243208.6   &                    & $17.5$ & 3  & 39       \\  
J102151.64+301011.9   &                    & $18.0$ & 12 & 34  & J170810.97+244341.6   &                    & $18.2$ & 3  & 16       \\  
J103549.68+092551.9   &                    & $16.0$ & 3  & 59  & J171617.33+553446.7   & SBSS\,1715+556     & $16.9$ & 8  & 39       \\  
J110215.97+521858.1   &                    & $17.2$ & 3  & 44  & J171629.92+575121.2   &                    & $17.9$ & 4  & 21       \\  
J110445.01+092530.9   &                    & $16.0$ & 4  & 40  & J172624.10+274419.3   & PG\,1724+278       & $15.7$ & 4  & 107       \\ 
J112242.69+613758.5   & PG\,1119+619       & $15.1$ & 3  & 87  & J174516.32+244348.3   &                    & $17.4$ & 3  & 22       \\  
J112414.45+402637.1   &                    & $17.7$ & 3  & 21  & J175125.67+255003.5   &                    & $17.2$ & 4  & 50       \\  
J113303.70+290223.0   &                    & $17.4$ & 3  & 34  & J202313.83+131254.9   &                    & $17.0$ & 3  & 33       \\  
J113418.00+015322.1   & LBQS\,1131+0209    & $17.7$ & 6  & 30  & J202758.63+773924.5   &                    & $17.7$ & 3  & 22       \\  
J113840.68$-$003531.7 & PG\,1136$-$003     & $14.2$ & 10 & 174 & J204300.90+002145.0   &                    & $17.6$ & 9  & 50       \\  
J113935.45+614953.9   & FBS\,1136+621      & $16.8$ & 3  & 34  & J204448.63+153638.8   &                    & $17.7$ & 7  & 50       \\  
J115358.81+353929.0   & FBS\,1151+359      & $16.3$ & 3  & 48  & J204546.81$-$054355.6 &                    & $17.8$ & 4  & 29       \\  
J115716.37+612410.7   & FBS\,1154+617      & $16.9$ & 5  & 34  & J204613.40$-$045418.7 &                    & $16.0$ & 3  & 120       \\ 
J125702.30+435245.8   &                    & $17.9$ & 3  & 18  & J204940.85+165003.6   &                    & $17.7$ & 7  & 35       \\  
J130059.20+005711.7   & PG\,1258+012       & $16.3$ & 3  & 47  & J210454.89+110645.5   &                    & $17.2$ & 4  & 37       \\  
J130439.57+312904.8   & LB\,28             & $16.8$ & 3  & 42  & J211651.96+003328.5   &                    & $17.7$ & 3  & 19       \\  
J133638.81+111949.4   &                    & $17.0$ & 3  & 32  & J215648.71+003620.7   & PB\,5010           & $17.7$ & 3  & 22       \\  
J134352.14+394008.3   &                    & $18.1$ & 3  & 19  & J225638.34+065651.1   & PG\,2254+067       & $15.1$ & 3  & 86       \\  
J135807.96+261215.5   &                    & $17.7$ & 4  & 23  & J232757.46+483755.2   &                    & $15.6$ & 3  & 92       \\  
J140545.25+014419.0   & PG\,1403+019       & $15.6$ & 3  & 81  & J233406.11+462249.3   &                    & $17.4$ & 3  & 35       \\  
J141549.05+111213.9   &                    & $15.8$ & 3  & 82  & J234528.85+393505.2   &                    & $17.3$ & 3  & 37       \\  
J143153.05$-$002824.3 & LBQS\,1429$-$0015  & $17.8$ & 8  & 34  & & & & &  \\
\noalign{\smallskip}
\hline
\end{tabular}
\end{center}
\end{table*}

\begin{table*}[t!]
\caption{Priority targets for follow-up (HRV subsample). $\dag$The binary system has been analysed in Geier et al. (\cite{geier11}).}
\label{targets:hrv}
\begin{center}
\begin{tabular}{lllllllll} \hline
\noalign{\smallskip}
Object            &  Class   & $T_{\rm eff}$  & $\log{g}$     &  $\log{y}$     &  $d$           &  $\Delta RV$          & $\Delta t$ \\ 
                  &          & [K]            &               &                &  [${\rm kpc}$] &  [${\rm km\,s^{-1}}$] & [d]        \\ \hline
\noalign{\smallskip}
J102151.64+301011.9    & sdB    & $30700\pm500$  & $5.71\pm0.06$ &  $<-3.0$      & $5.8_{-0.5}^{+0.5}$ & $277\pm51$           &  $14.936$    \\
J150829.02+494050.9    & sdB    & $28200\pm600$  & $5.34\pm0.09$ &  $-2.0\pm0.2$ & $6.4_{-0.7}^{+0.8}$  & $211\pm18$           &  $2161.429$  \\
J095229.62+301553.6    & sdB    & $35200\pm1200$ & $5.05\pm0.17$ &  $<-3.0$      & $16.0_{-3.3}^{+3.8}$  & $198\pm40$           &  $1155.766$  \\
J113840.68$-$003531.7$\dag$  & sdB    & $30800\pm500$  & $5.50\pm0.09$ &  $-3.0\pm0.2$ & $1.3_{-0.1}^{+0.2}$  & $182\pm12$           &  $0.973$     \\ 
J165404.26+303701.8$\dag$    & sdB    & $24400\pm800$  & $5.32\pm0.11$ &  $-2.3\pm0.3$ & $1.9_{-0.3}^{+0.3}$  & $181\pm9$            &  $1795.144$ \\ 
J152222.15$-$013018.3  & sdB    & $24800\pm1000$ & $5.52\pm0.15$ &  $-2.6\pm0.5$ & $4.8_{-0.9}^{+1.1}$  & $173\pm36$           &  $3.001$     \\ 
J150513.52+110836.6$\dag$    & sdB    & $33300\pm500$  & $5.80\pm0.10$ &  $-2.4\pm0.3$ & $1.5_{-0.2}^{+0.2}$  &   $154\pm12$           &  $0.957$     \\ 
J002323.99$-$002953.2$\dag$  & sdB    & $30100\pm500$  & $5.62\pm0.08$ &  $-2.2\pm0.2$ & $1.8_{-0.2}^{+0.2}$  &   $130\pm14$           &  $82.784$    \\ 
J202313.83+131254.9    & sdB    & $29600\pm600$  & $5.64\pm0.14$ &  $-2.1\pm0.1$ & $3.8_{-0.6}^{+0.7}$  &   $124\pm21$           &  $1202.795$   \\ 
J012022.94+395059.4    & sdB    & $28900\pm500$  & $5.51\pm0.08$ &  $-3.0\pm0.4$ & $1.9_{-0.2}^{+0.2}$  &   $114\pm11$           &  $360.973$   \\ 
J202758.63+773924.5    & sdO    & $46200\pm3200$ & $5.48\pm0.18$ &  $-2.8\pm0.9$ & $8.2_{-1.8}^{+2.2}$  & $114\pm48$           &  $1.960$   \\ 
J095238.93+252159.7    & sdB    & $27800\pm500$  & $5.61\pm0.08$ &  $-2.64\pm0.1$ & $1.2_{-0.1}^{+0.1}$ &   $111\pm10$           &  $2.918$   \\ 
J161140.50+201857.0    & sdOB   & $36900\pm700$  & $5.89\pm0.13$ &  $-1.2\pm0.1$ & $6.1_{-0.9}^{+1.1}$  & $108\pm36$           &  $0.947$   \\ 
J164326.04+330113.1    & sdB    & $27900\pm500$  & $5.62\pm0.07$ &  $-2.3\pm0.2$ & $2.4_{-0.2}^{+0.2}$  &   $108\pm11$           &  $1.990$   \\
J204448.63+153638.8    & sdB    & $29600\pm600$  & $5.57\pm0.09$ &  $-2.2\pm0.1$ & $5.7_{-0.7}^{+0.7}$  & $101\pm19$           &  $3.049$   \\
J083006.17+475150.4    & sdB    & $25200\pm500$  & $5.30\pm0.05$ &  $-3.3\pm0.7$ & $2.8_{-0.2}^{+0.2}$  &   $95\pm14$            &  $3.961$   \\ 
J204940.85+165003.6    & He-sdO & $43000\pm700$  & $5.71\pm0.13$ &  $>+2.0$       & $6.2_{-0.9}^{+1.1}$ & $85\pm19$            &  $5.932$   \\
\noalign{\smallskip}
\hline
\end{tabular}                                                                           
\end{center}
\end{table*}

\begin{table*}[t!]
\caption{Priority targets for follow-up (RRV subsample). $\dag$The binary system has been analysed in Geier et al. (\cite{geier11}). $\ddag$Atmospheric parameters ($T_{\rm eff}=39400\,{\rm K}$, $\log{g}=5.64$, $\log{y}=-0.55$) have been determined by Str\"oer et al. (\cite{stroeer07}).}
\label{targets:rrv}
\begin{center}
\begin{tabular}{lllllllll} \hline
\noalign{\smallskip}
J085727.65+424215.4    & He-sdO & $39500\pm1900$ & $5.63\pm0.24$ &  $+0.2\pm0.2$ & $8.7_{-2.2}^{+3.0}$  &  $111\pm46$           &  $0.066$   \\
J161817.65+120159.6    & sdB    & $32100\pm1000$ & $5.35\pm0.23$ &  $-$          & $8.1_{-2.1}^{+2.8}$  & $105\pm31$           &  $0.043$   \\
J232757.46+483755.2    & He-sdO & $64700\pm2000$ & $5.40\pm0.08$ &  $>+2.0$      & $4.2_{-0.4}^{+0.5}$  & $105\pm24$           &  $0.016$   \\
J162256.66+473051.1    & sdB    & $28600\pm500$  & $5.70\pm0.11$ &  $-1.81\pm0.1$ & $2.2_{-0.3}^{+0.3}$ &   $101\pm15$           &  $0.037$   \\
J163702.78$-$011351.7  & He-sdO & $46100\pm700$  & $5.92\pm0.22$ &  $>+2.0$      & $3.8_{-0.9}^{+1.1}$ &   $101\pm55$           &  $0.085$   \\
J113303.70+290223.0    & sdB/DA & $-$            & $-$           &  $-$           &      $-$           &  $95\pm35$            &  $0.016$   \\
J135807.96+261215.5    & sdB    & $33500\pm600$  & $5.66\pm0.10$ &  $<-2.0$       & $5.8_{-0.7}^{+0.8}$ & $87\pm29$            &  $0.030$   \\
J112242.69+613758.5    & sdB    & $29300\pm500$  & $5.69\pm0.10$ &  $-2.3\pm0.3$  & $1.5_{-0.2}^{+0.2}$ &  $83\pm20$            &  $0.047$   \\
J153411.10+543345.2    & sdOB   & $34800\pm700$  & $5.64\pm0.09$ &  $-2.6\pm0.3$  & $3.8_{-0.4}^{+0.5}$ &  $83\pm29$            &  $0.018$   \\
J082053.53+000843.4    & sdB    & $26700\pm900$  & $5.48\pm0.10$ &  $-2.0\pm0.09$ & $1.6_{-0.2}^{+0.3}$ &  $77\pm11$            &  $0.047$   \\
J170810.97+244341.6    & sdOB   & $35600\pm800$  & $5.58\pm0.14$ &  $-0.8\pm0.1$  & $8.5_{-1.4}^{+1.6}$ &  $76\pm33$            &  $0.013$   \\
J094856.95+334151.0    & He-sdO & $51000\pm1200$ & $5.87\pm0.12$ &  $+1.8\pm0.5$  & $5.1_{-0.7}^{+0.8}$ &  $75\pm17$            &  $0.012$   \\
J204613.40$-$045418.7$\dag$  & sdB    & $31600\pm600$  & $5.55\pm0.10$ &  $-3.7\pm0.6$  & $2.8_{-0.4}^{+0.4}$ &  $70\pm13$            &  $0.030$   \\
J215648.71+003620.7    & sdB    & $30800\pm800$  & $5.77\pm0.12$ &  $-2.2\pm0.3$  & $4.7_{-0.7}^{+0.8}$ &  $69\pm21$            &  $0.011$   \\
J074534.16+372718.5    & sdB    & $37500\pm500$  & $5.90\pm0.09$ &  $<-3.0$       & $4.6_{-0.5}^{+0.5}$ &  $65\pm19$            &  $0.036$   \\
J143153.05$-$002824.3  & sdOB   & $37300\pm800$  & $6.02\pm0.16$ &  $-0.8\pm0.1$  & $4.4_{-0.8}^{+0.9}$ &  $65\pm22$            &  $0.012$   \\
J171629.92+575121.2    & sdOB   & $35400\pm1000$ & $5.60\pm0.18$ &  $-0.7\pm0.1$  & $7.8_{-0.9}^{+1.0}$ &  $65\pm16$            &  $0.013$   \\
J112414.45+402637.1    & He-sdO & $47100\pm1000$ & $5.81\pm0.23$ &  $+1.7\pm0.7$  & $5.9_{-1.4}^{+1.9}$ &  $63\pm22$            &  $0.021$   \\
J125702.30+435245.8    & sdB    & $28000\pm1100$ & $5.77\pm0.17$ &  $<-3.0$       & $4.9_{-1.0}^{+1.3}$ &  $63\pm28$            &  $0.010$   \\
J110215.97+521858.1    & He-sdO & $56600\pm4200$ & $5.36\pm0.22$ &  $>+2.0$       & $8.9_{-2.2}^{+3.0}$ &  $62\pm11$            &  $0.033$   \\
J151415.66$-$012925.2  & He-sdO & $48200\pm500$  & $5.85\pm0.08$ &  $+1.7\pm0.4$  & $3.6_{-0.3}^{+0.4}$ &  $62\pm22$            &  $0.016$   \\
J204300.90+002145.0    & sdO    & $40200\pm700$  & $6.15\pm0.13$ &  $-1.3\pm0.4$  & $3.6_{-0.5}^{+0.6}$ &  $61\pm13$            &  $0.016$   \\ 
J171617.33+553446.7    & sdB    & $32900\pm900$  & $5.48\pm0.09$ &  $<-3.0$       & $4.9_{-0.6}^{+0.7}$ & $60\pm24$           &  $0.048$   \\
J210454.89+110645.5    & sdOB   & $37800\pm700$  & $5.63\pm0.10$ &  $-2.4\pm0.2$  & $4.9_{-0.6}^{+0.6}$ &  $58\pm19$            &  $0.023$   \\ 
J115358.81+353929.0    & sdOB   & $29400\pm500$  & $5.49\pm0.06$ &  $-2.5\pm0.3$  & $3.3_{-0.3}^{+0.3}$ &  $56\pm12$            &  $0.022$   \\ 
J174516.32+244348.3    & He-sdO & $43400\pm1000$ & $5.62\pm0.21$ &  $>+2.0$       & $6.2_{-1.4}^{+1.8}$ &  $55\pm28$            &  $0.016$   \\ 
J134352.14+394008.3    & He-sdB & $36000\pm2100$ & $4.78\pm0.30$ &  $-0.2\pm0.2$  & $8.8_{-6.1}^{+8.5}$ &  $52\pm34$            &  $0.022$   \\ 
J115716.37+612410.7    & sdB    & $29900\pm500$  & $5.59\pm0.08$ &  $-3.2\pm0.8$  & $4.0_{-0.4}^{+0.5}$ &  $51\pm34$            &  $0.049$   \\ 
J133638.81+111949.4    & sdB    & $27500\pm500$  & $5.49\pm0.08$ &  $-2.7\pm0.2$  & $4.4_{-0.5}^{+0.5}$ &  $48\pm17$            &  $0.030$   \\ 
J211651.96+003328.5    & sdB    & $27900\pm800$  & $5.78\pm0.15$ &  $-3.9\pm0.7$  & $4.3_{-0.8}^{+0.9}$ &  $48\pm23$            &  $0.016$   \\ 
J170645.57+243208.6    & sdB    & $32000\pm500$  & $5.59\pm0.07$ &  $<-4.0$       & $5.5_{-0.5}^{+0.6}$ &  $46\pm14$            &  $0.013$   \\ 
J175125.67+255003.5    & sdB    & $30600\pm500$  & $5.48\pm0.08$ &  $<-3.8$       & $5.0_{-0.5}^{+0.6}$ &  $46\pm14$            &  $0.034$   \\ 
J012739.35+404357.8    & sdO    & $48300\pm3200$ & $5.67\pm0.10$ &  $-1.3\pm0.2$  & $4.1_{-0.6}^{+0.7}$ &  $45\pm17$            &  $0.037$   \\ 
J113418.00+015322.1    & sdB    & $29700\pm1200$ & $4.83\pm0.16$ & $<-4.0$        & $1.8_{-2.4}^{+2.9}$ &  $45\pm24$            &  $0.076$   \\ 
J172624.10+274419.3$\dag$    & sdOB   & $33500\pm500$  & $5.71\pm0.09$ &  $-2.2\pm0.1$  & $2.2_{-0.2}^{+0.3}$ &  $45\pm16$            &  $0.047$   \\ 
J155628.34+011335.0    & sdB    & $32700\pm600$  & $5.51\pm0.08$ &  $-2.9\pm0.2$  & $3.1_{-0.3}^{+0.4}$ &  $44\pm15$            &  $0.068$   \\ 
J103549.68+092551.9    & He-sdO & $48100\pm600$  & $6.02\pm0.13$ &  $>+2.0$       & $2.2_{-0.3}^{+0.4}$ &  $43\pm12$            &  $0.021$   \\ 
J141549.05+111213.9    & He-sdO & $43100\pm800$  & $5.81\pm0.17$ &  $>+2.0$       & $2.4_{-0.4}^{+0.5}$ &  $43\pm7$             &  $0.023$   \\ 
J152705.03+110843.9    & sdOB   & $37600\pm500$  & $5.62\pm0.10$ &  $-0.5\pm0.1$  & $4.8_{-0.5}^{+0.6}$ &  $43\pm14$            &  $0.054$   \\ 
J052544.93+630726.0    & sdOB   & $35600\pm800$  & $5.85\pm0.10$ &  $-1.6\pm0.2$  & $4.3_{-0.5}^{+0.6}$ &  $42\pm17$            &  $0.026$   \\ 
J100535.76+223952.1    & sdB    & $29000\pm700$  & $5.43\pm0.13$ &  $-2.7\pm0.2$  & $7.9_{-1.3}^{+1.5}$ &  $41\pm18$            &  $0.019$   \\ 
J204546.81$-$054355.6  & sdB    & $35500\pm500$  & $5.47\pm0.09$ &  $-1.4\pm0.2$  & $7.3_{-0.8}^{+0.9}$ &  $41\pm18$            &  $0.013$   \\ 
J092520.70+470330.6    & sdB    & $28100\pm900$  & $5.17\pm0.15$ &  $-2.5\pm0.2$  & $7.5_{-1.4}^{+1.7}$ &  $40\pm13$            &  $0.012$   \\ 
J075937.15+541022.2    & sdB    & $31300\pm700$  & $5.30\pm0.10$ &  $-3.3\pm0.3$  & $7.6_{-1.0}^{+1.1}$ &  $38\pm13$            &  $0.012$   \\ 
J234528.85+393505.2    & He-sdO & $47900\pm800$  & $6.07\pm0.14$ &  $>+2.0$       & $3.5_{-0.5}^{+0.6}$ &  $37\pm14$            &  $0.012$   \\
J130439.57+312904.8    & sdOB   & $38100\pm600$  & $5.69\pm0.12$ &  $-0.4\pm0.1$  & $4.1_{-0.6}^{+0.6}$ &  $36\pm12$            &  $0.037$   \\ 
J130059.20+005711.7$\ddag$    & He-sdO & $40700\pm500$  & $5.53\pm0.10$ &  $-0.6\pm0.1$  & $3.9_{-0.4}^{+0.5}$ &  $36\pm16$            &  $0.012$   \\ 
J110445.01+092530.9    & sdOB   & $35900\pm800$  & $5.41\pm0.07$ &  $-2.1\pm0.4$  & $3.8_{-0.3}^{+0.4}$ &  $34\pm14$            &  $0.040$   \\ 
J113935.45+614953.9    & sdB    & $28800\pm900$  & $5.27\pm0.15$ &  $-2.8\pm0.3$  & $4.9_{-0.9}^{+1.1}$ &  $31\pm14$            &  $0.011$   \\ 
J233406.11+462249.3    & sdOB   & $34600\pm500$  & $5.71\pm0.09$ &  $-1.3\pm0.1$  & $4.9_{-0.6}^{+0.6}$ &  $31\pm14$            &  $0.025$   \\ 
J225638.34+065651.1$\dag$    & sdB    & $28900\pm600$  & $5.58\pm0.11$ &  $-3.0\pm0.2$  & $1.6_{-0.2}^{+0.3}$ &  $27\pm11$            &  $0.031$   \\ 
J140545.25+014419.0    & sdB    & $27300\pm800$  & $5.37\pm0.16$ &  $-1.9\pm0.2$  & $2.5_{-0.5}^{+0.6}$ &  $25\pm10$            &  $0.026$   \\
\noalign{\smallskip}
\hline                                                                              
\end{tabular}                                                                           
\end{center}
\end{table*}

\begin{appendix}
\onecolumn
\section{Close binary subdwarfs from literature}
\begin{table*}[t!]
\caption{Orbital parameters of all known hot subdwarf binaries from literature. The superscript p denotes sdB pulsators, r binaries where with reflection effect, ec eclipsing systems and el systems with light variations caused by ellipsoidal deformation.\dag Post-RGB stars without core helium-burning. \ddag Double-lined binary consisting of two helium rich sdBs. The RV semi-amplitudes of both components are given.} 
\label{tab:orbitslit}
\begin{center}
\begin{tabular}{llrrl} 
Object & P &  $\gamma$ & K & Reference\\
 & [d] & [${\rm km\,s^{-1}}$] & [${\rm km\,s^{-1}}$] &\\ 
\noalign{\smallskip}
\hline
\noalign{\smallskip}
PG\,0850+170     &  $27.815$        &  $32.2\pm2.8$  &  $33.5\pm3.3$ &  Morales-Rueda et al. \cite{morales03a} \\
PG\,1619+522     &  $15.3578$       &  $-52.5\pm1.1$ &  $35.2\pm1.1$ &  Morales-Rueda et al. \cite{morales03a} \\
PG\,1110+294     &  $9.4152$        &  $-15.2\pm0.9$ &  $58.7\pm1.2$ &  Morales-Rueda et al. \cite{morales03a} \\
Feige\,108       &  $8.7465$        &  $45.8\pm0.6$  &  $50.2\pm1.0$ &  Edelmann et al. \cite{edelmann04} \\
PG\,0940+068     &  $8.330$         &  $-16.7\pm1.4$ &  $61.2\pm1.4$ &  Maxted et al. \cite{maxted00b} \\
PHL\,861         &  $7.44$          &  $-26.5\pm0.4$ &  $47.9\pm0.4$ &  Karl et al. \cite{karl06} \\
HE\,1448$-$0510  &  $7.159$         &  $-45.5\pm0.8$ &  $53.7\pm1.1$ &  Karl et al. \cite{karl06} \\
PG\,1032+406     &  $6.7791$        &  $24.5\pm0.5$  &  $33.7\pm0.5$ &  Morales-Rueda et al. \cite{morales03a} \\
PG\,0907+123     &  $6.11636$       &  $56.3\pm1.1$  &  $59.8\pm0.9$ &  Morales-Rueda et al. \cite{morales03a} \\
HE\,1115$-$0631  &  $5.87$          &  $87.1\pm1.3$  &  $61.9\pm1.1$ &  Napiwotzki et al. in prep. \\
CD\,$-$24\,731   &  $5.85$          &  $20.0\pm5.0$  &  $63.0\pm3.0$ &  Edelmann et al. \cite{edelmann05} \\
PG\,1244+113     &  $5.75207$       &  $9.8\pm1.2$   &  $55.6\pm1.8$ &  Morales-Rueda et al. \cite{morales03b} \\
PG\,0839+399     &  $5.6222$        &  $23.2\pm1.1$  &  $33.6\pm1.5$ &  Morales-Rueda et al. \cite{morales03a} \\
TON\,S\,135      &  $4.1228$        &  $-3.7\pm1.1$  &  $41.4\pm1.5$ &  Edelmann et al. \cite{edelmann05} \\
PG\,0934+186     &  $4.051$         &  $7.4\pm2.9$   &  $60.2\pm2.0$ &  Morales-Rueda et al. \cite{morales03b} \\
PB\,7352         &  $3.62166$       &  $-2.1\pm0.3$  &  $60.8\pm0.3$ &  Edelmann et al. \cite{edelmann05} \\
KPD\,0025+5402   &  $3.5711$        &  $-7.8\pm0.7$  &  $40.2\pm1.1$ &  Morales-Rueda et al. \cite{morales03a} \\
TON\,245         &  $2.501$         &  $-$           &  $88.3$       &  Morales-Rueda et al. \cite{morales03a} \\
PG\,1300+2756    &  $2.25931$       &  $-3.1\pm0.9$  &  $62.8\pm1.6$ &  Morales-Rueda et al. \cite{morales03a} \\ 
NGC\,188/II$-$91 &  $2.15$          &  $-$           &  $22.0$       &  Green et al. \cite{green04} \\ 
V\,1093\,Her$^{\rm p}$   &  $1.77732$       &  $-3.9\pm0.8$  &  $70.8\pm1.0$ &  Morales-Rueda et al. \cite{morales03a} \\
HD\,171858       &  $1.63280$       &  $62.5\pm0.1$  &  $60.8\pm0.3$ &  Edelmann et al. \cite{edelmann05} \\
KPD\,2040+3954   &  $1.48291$       &  $-11.5\pm1.0$ &  $95.1\pm1.7$ &  Morales-Rueda et al. \cite{morales03b} \\
HE\,2150$-$0238  &  $1.321$         &  $-32.5\pm0.9$ &  $96.3\pm1.4$ &  Karl et al. \cite{karl06} \\
$[$CW83$]$\,1735+22  &  $1.278$         &  $20.6\pm0.4$  &  $103.0\pm1.5$ &  Edelmann et al. \cite{edelmann05} \\
PG\,1512+244     &  $1.26978$       &  $-2.9\pm1.0$  &  $92.7\pm1.5$ & Morales-Rueda et al. \cite{morales03a} \\
PG\,0133+114     &  $1.23787$       &  $-0.3\pm0.2$  &  $82.0\pm0.3$ & Edelmann et al. \cite{edelmann05} \\
HE\,1047$-$0436  &  $1.21325$       &  $25.0\pm3.0$  &  $94.0\pm3.0$ & Napiwotzki et al. \cite{napiwotzki01} \\
HE\,1421$-$1206  &  $1.188$         &  $-86.2\pm1.1$ &  $55.5\pm2.0$ & Napiwotzki et al. in prep. \\
PG\,1000+408     &  $1.041145$      &  $41.9$        &  $72.4$       & Shimanskii et al. \cite{shimanskii08} \\
PB\,5333         &  $0.92560$       &  $-95.3\pm1.3$ &  $22.4\pm0.8$ & Edelmann et al. \cite{edelmann04} \\
HE\,2135$-$3749  &  $0.9240$        &  $45.0\pm0.5$  &  $90.5\pm0.6$ & Karl et al. \cite{karl06} \\
EC\,12408$-$1427 &  $0.90243$       &  $-52.0\pm1.2$ &  $58.9\pm1.6$ & Morales-Rueda et al. \cite{morales06} \\
PG\,0918+0258    &  $0.87679$       &  $104.4\pm1.7$ &  $80.0\pm2.6$ & Morales-Rueda et al. \cite{morales03a} \\
PG\,1116+301     &  $0.85621$       &  $-0.2\pm1.1$  &  $88.5\pm2.1$ & Morales-Rueda et al. \cite{morales03a} \\
PG\,1230+052     &  $0.8372$        &  $-43.4\pm0.8$ &  $41.5\pm1.3$ & Morales-Rueda et al. \cite{morales03b} \\
V\,2579\,Oph$^{\rm p}$     &  $0.8292056$     &  $-54.16\pm0.27$ &  $70.10\pm0.13$ & For et al. \cite{for06} \\
TON\,S\,183      &  $0.8277$        &  $50.5\pm0.8$  &  $84.8\pm1.0$ & Edelmann et al. \cite{edelmann05} \\
EC\,02200$-$2338 &  $0.8022$        &  $20.7\pm2.3$  &  $96.3\pm1.4$ & Morales-Rueda et al. \cite{morales05} \\
PG\,0849+319     &  $0.74507$       &  $64.0\pm1.5$  &  $66.3\pm2.1$ & Morales-Rueda et al. \cite{morales03a} \\
JL\,82$^{\rm r}$ &  $0.73710$       &  $-1.6\pm0.8$  &  $34.6\pm1.0$ & Edelmann et al. \cite{edelmann05} \\
PG\,1248+164     &  $0.73232$       &  $-16.2\pm1.3$ &  $61.8\pm1.1$ & Morales-Rueda et al. \cite{morales03a} \\
HD\,188112\dag   &  $0.60658125$    &  $26.6\pm0.3$  &  $188.4\pm0.2$ & Edelmann et al. \cite{edelmann05} \\
PG\,1247+554     &  $0.602740$      &  $13.8\pm0.6$  &  $32.2\pm1.0$ & Maxted et al. \cite{maxted00b} \\
PG\,1725+252     &  $0.601507$      &  $-60.0\pm0.6$ &  $104.5\pm0.7$ & Morales-Rueda et al. \cite{morales03a} \\
PG\,0101+039$^{\rm el,p}$   &  $0.569899$      &  $7.3\pm0.2$ &  $104.7\pm0.4$ & Geier et al. \cite{geier08} \\
HE\,1059$-$2735  &  $0.555624$      &  $-44.7\pm0.6$ &  $87.7\pm0.8$ & Napiwotzki et al. in prep. \\
PG\,1519+640     &  $0.54029143$    &  $0.1\pm0.4$   & $42.7\pm0.6$ & Edelmann et al. \cite{edelmann04} \\
PG\,0001+275     &  $0.529842$      &  $-44.7\pm0.5$ & $92.8\pm0.7$ & Edelmann et al. \cite{edelmann05} \\
PG\,1743+477     &  $0.515561$      &  $-65.8\pm0.8$ & $121.4\pm1.0$ & Morales-Rueda et al. \cite{morales03a} \\
HE\,1318$-$2111  &  $0.487502$      &  $48.9\pm0.7$  & $48.5\pm1.2$ & Napiwotzki et al. in prep. \\
PG\,1544$+$488\ddag  &  $0.48$          &  $-23\pm4$     & $57\pm4/97\pm10$ & Ahmad et al. \cite{ahmad04} \\
GALEX\,J234947.7$+$384440 & $0.46249$ & $2.0\pm1.0$  & $87.9\pm2.2$ & Kawka et al. \cite{kawka10} \\
HE\,0230$-$4323$^{\rm r,p}$   &  $0.45152$       &  $16.6\pm1.0$ &  $62.4\pm1.6$ & Edelmann et al. \cite{edelmann05} \\
HE\,0929$-$0424  &  $0.4400$        &  $41.4\pm1.0$  & $114.3\pm1.4$ & Karl et al. \cite{karl06} \\
\noalign{\smallskip}
\hline 
\end{tabular}
\end{center}
\end{table*}

\begin{table*}[t!]
\begin{center}
\begin{tabular}{llrrl} 
Object & P &  $\gamma$ & K & Reference\\
 & [d] & [${\rm km\,s^{-1}}$] & [${\rm km\,s^{-1}}$] &\\
\noalign{\smallskip} 
\hline
\noalign{\smallskip}
$[$CW83$]$\,1419$-$09 & $0.4178$        &  $42.3\pm0.3$  & $109.6\pm0.4$ & Edelmann et al. \cite{edelmann05} \\
KPD\,1946+4340$^{\rm ec,el}$   &  $0.403739$      &  $-5.5\pm1.0$  & $167.0\pm2.4$ & Morales-Rueda et al. \cite{morales03a} \\
KUV\,04421+1416$^{\rm r,p}$           &  $0.398$         &  $33\pm3$   & $90\pm5$ & Reed et al. \cite{reed10} \\
Feige\,48$^{\rm p}$        &  $0.376$         &  $-47.9\pm0.1$ & $28.0\pm0.2$  & O'Toole et al. \cite{otoole04} \\
GD\,687          &  $0.37765$       &  $32.3\pm3.0$  & $118.3\pm3.4$ & Geier et al. \cite{geier10a} \\
PG\,1232$-$136   &  $0.3630$        &  $4.1\pm0.3$   & $129.6\pm0.04$ & Edelmann et al. \cite{edelmann05} \\
PG\,1101+249     &  $0.35386$       &  $-0.8\pm0.9$  & $134.6\pm1.3$ & Moran et al. \cite{moran99} \\
PG\,1438$-$029$^{\rm r}$   &  $0.336$         &  $-$           & $32.1$        & Green et al. \cite{green05} \\
PG\,1528+104     &  $0.331$         &  $-49.9\pm0.8$ & $52.7\pm1.3$ & Morales-Rueda et al. \cite{morales03b} \\
PG\,0941+280$^{\rm ec}$     &  $0.315$         &  $-$           & $-$          & Green et al. \cite{green04} \\
KBS\,13$^{\rm r}$ &  $0.2923$        &  $7.53\pm0.08$ & $22.82\pm0.23$ & For et al. \cite{for08} \\
CPD$-$64\,481    &  $0.2772$        &  $94.1\pm0.3$  & $23.8\pm0.4$ & Edelmann et al. \cite{edelmann05} \\
GALEX\,J032139.8$+$472716 & $0.26584$  &  $70.5\pm2.2$  &  $59.8\pm 4.5$ & Kawka et al. \cite{kawka10} \\
HE\,0532$-$4503  &  $0.2656$        &  $8.5\pm0.1$   & $101.5\pm0.2$ & Karl et al. \cite{karl06} \\
AA\,Dor$^{\rm ec,r}$       &  $0.2614$        &  $1.57\pm0.09$ & $40.15\pm0.11$ & M\"uller et al. \cite{mueller10} \\
PG\,1329+159$^{\rm r}$     &  $0.249699$      &  $-22.0\pm1.2$ & $40.2\pm1.1$ & Morales-Rueda et al. \cite{morales03a} \\
PG\,2345+318$^{\rm ec}$    &  $0.2409458$     &  $-10.6\pm1.4$ & $141.2\pm1.1$ & Moran et al. \cite{moran99} \\
PG\,1432+159               &  $0.22489$       &  $-16.0\pm1.1$ & $120.0\pm1.4$ & Moran et al. \cite{moran99} \\
BPS\,CS\,22169$-$0001$^{\rm r}$      &  $0.1780$        &  $2.8\pm0.3$   & $14.9\pm0.4$  & Edelmann et al. \cite{edelmann05} \\
HS\,2333+3927$^{\rm r}$    &  $0.1718023$     &  $-31.4\pm2.1$ & $89.6\pm3.2$ & Heber et al. \cite{heber04} \\
2M\,1533+3759$^{\rm ec,r}$ &  $0.16177042$    &  $-3.4\pm5.2$  & $71.1\pm1.0$ & For et al. \cite{for10} \\
EC\,00404$-$4429           &  $0.12834$       &  $33.0\pm2.9$  & $152.8\pm3.4$ & Morales-Rueda et al. \cite{morales05} \\
2M\,1938+4603$^{\rm ec,r}$ &  $0.1257653$     &  $20.1\pm0.3$  & $65.7\pm0.6$  & \O stensen et al. \cite{oestensen10} \\
BUL-SC\,16\,335$^{\rm ec,r}$            &  $0.125050278$   &  $-$           & $-$          & Polubek et al. \cite{polubek07} \\
PG\,1043+760               &  $0.1201506$     &  $24.8\pm1.4$  & $63.6\pm1.4$ &  Morales-Rueda et al. \cite{morales03a} \\
HW\,Vir$^{\rm ec,r}$       &  $0.115$         &  $-13.0\pm0.8$ & $84.6\pm1.1$ &  Edelmann \cite{edelmann08} \\
HS\,2231+2441$^{\rm ec,r}$ &  $0.1105880$     &  $-$           & $49.1\pm3.2$ &  \O stensen et al. \cite{oestensen07} \\
NSVS\,14256825$^{\rm ec,r}$ & $0.110374102$   &  $-$           & $-$          &  Wils et al. \cite{wils07} \\
PG\,1336$-$018$^{\rm ec,r,p}$  &  $0.101015999$   &  $-25.0$       & $78.7\pm0.6$ &  Vu\v ckovi\'c et al. \cite{vuckovic07} \\
HS\,0705+6700$^{\rm ec,r}$ &  $0.09564665$    &  $-36.4\pm2.9$ & $85.8\pm3.6$ &  Drechsel et al. \cite{drechsel01} \\
KPD\,1930+2752$^{\rm el,p}$ & $0.0950933$     &  $5.0\pm1.0$   & $341.0\pm1.0$ &  Geier et al. \cite{geier07} \\
KPD\,0422+5421$^{\rm ec,el}$ & $0.09017945$   &  $-57.0\pm12.0$ & $237.0\pm18.0$ & Orosz \& Wade \cite{orosz99} \\
NGC\,6121$-$V46$^{\rm el}$\dag &  $0.087159$      &  $31.3\pm1.6$  & $211.6\pm2.3$ & O'Toole et al. \cite{otoole06} \\
PG\,1017$-$086$^{\rm r}$   &  $0.0729938$     &  $-9.1\pm1.3$  & $51.0\pm1.7$  & Maxted et al. \cite{maxted02} \\
\noalign{\smallskip}
\hline
\end{tabular}
\end{center}
\end{table*}
\end{appendix}

\end{document}